\documentclass{jfm}

\usepackage{amsmath,amssymb,mathtools,bm}
\usepackage{multirow}
\usepackage{graphicx}
\usepackage{overpic}
\usepackage{color}
\usepackage{tabularx}
\usepackage{siunitx}
\usepackage[normalem]{ulem}
\newcolumntype{Y}{X}
\usepackage{tabto}
\usepackage{nicematrix}

\usepackage{array}
\newcolumntype{L}[1]{>{\raggedright\let\newline\\\arraybackslash\hspace{0pt}}m{#1}}
\newcolumntype{C}[1]{>{\centering\let\newline\\\arraybackslash\hspace{0pt}}m{#1}}
\newcolumntype{R}[1]{>{\raggedleft\let\newline\\\arraybackslash\hspace{0pt}}m{#1}}

\usepackage{relsize,exscale}
\usepackage[skip=10pt plus1pt, indent=10pt]{parskip}
\usepackage{natbib} 
\usepackage{my_macros}

\renewcommand{\vec}[1]{\bm{#1}}
\newcommand{\tensor}[1]{\mathsfbi{#1}}
\newcommand{\mat}[1]{\tensor{#1}}
\newcommand{\intd}{\mathop{}\!\mathrm{d}}

\newcommand{\diff}[2]{\dfrac{\mathrm{d}{#1}}{\mathrm{d}{#2}}}
\newcommand{\pdiff}[2]{\dfrac{\partial{#1}}{\partial{#2}}}
\newcommand{\transpose}{^\top}
\newcommand{\bigO}[1]{\mathit{O}\left(#1\right)}

\newcommand{\ord}[1]{\mathrm{ord}\left(#1\right)}
\newcommand{\avg}[1]{\overline{#1}}

\newcommand{\e}{\vec{e}}

\newcommand{\x}{\vec{x}}
\newcommand{\vnabla}{\vec{\nabla}}

\newcommand{\flowVel}{\vec{u}^*}
\newcommand{\flowAngVel}{\vec{\Omega}^*}
\newcommand{\flowAngSpeed}{\Omega^*}
\newcommand{\flowStrainRate}{\tensor{E}^*}

\newcommand{\gait}{U}

\newcommand{\shearrate}{\gamma}

\newcommand{\UI}{\gait_I}
\newcommand{\B}{B}

\newcommand{\ybar}{\bar{y}_0}
\newcommand{\thetabar}{\bar{\theta}_0}

\newcommand{\inprod}[2]{\left\langle #1, #2 \right\rangle}

\usepackage[hidelinks,hypertexnames=false]{hyperref}
\usepackage[capitalise,nameinlink]{cleveref}
\AddToHook{cmd/appendix/before}{\crefalias{section}{appendix}}

\title{Motility and rotation of multi-timescale microswimmers in linear background flows}
\shorttitle{Motility and rotation of multi-timescale microswimmers in flow}
\shortauthor{E. A. Gaffney, K. Ishimoto and B. J. Walker}

\author{Eamonn A. Gaffney\aff{1}, Kenta Ishimoto\aff{2}, \and Benjamin J. Walker\aff{3\corresp{\email{benjamin.walker@ucl.ac.uk}}}}

\affiliation{\aff{1} Wolfson Centre for Mathematical Biology, Mathematical Institute, University of Oxford, Oxford, OX2 6GG, UK
\aff{2} Department of Mathematics, Kyoto University, Kyoto, 606-8502, Japan
\aff{3}Department of Mathematics, University College London, London, WC1H 0AY, UK}

\begin{document}

\maketitle 

\begin{abstract}
Microswimming cells and robots exhibit diverse behaviours due to both their swimming and their environment. One of the core environmental features impacting inertialess swimming is background flows. While the influence of select flows, particularly shear flows, have been extensively investigated, these are special cases. Here, we examine inertialess swimmers in more general flows, specifically general linear planar flows that may also possess rapid oscillations. Relatively weak symmetry constraints are imposed on the swimmer to ensure planarity and to reduce complexity. A further constraint reflecting common observation is imposed, namely that the swimmer is inefficient, which we suitably define. This introduces two separate timescales: a fast timescale associated with swimmer actuation, and a second timescale associated with net swimmer movement, with inefficiency dictating that this latter timescale is much slower, allowing for a multiple timescale simplification of the governing equations. With the exception of mathematically precise edge cases, we find that the behaviour of the swimmer is dictated by two parameter groupings, both of which measure balances between the angular velocity and rate of strain of the background flow. While the measures of flow angular velocity and strain rates that primarily govern the rotational dynamics are modulated by swimmer properties, the primary features of the translational motion are determined solely by a ratio of flow angular velocity to strain rate. Hence, a simple classification of the swimmer dynamics emerges. For example, this illustrates the limited extent to which, and how,  microswimmers may control their orientations and trajectories in flows.
\end{abstract}

\section{Introduction}

Microswimming cells, together with robotic swimmers at low Reynolds number, exhibit a myriad of behaviours and characteristics, in part due to the complexity of their actuation and in part due to the diversity of their surrounding fluid environment \citep{Lauga2009,Gaffney2011,Elgeti2015,Goldstein2015,diaz2021,Huang2016}. 
A common feature of the microswimming environment is the presence of a background flow, which can influence microswimming in diverse ways. For instance, flows can induce guidance cues for cell navigation, often referred to as rheotaxis. Examples include sperm motility \citep{Miki2013} and the behaviour of swimmers in microdevices, including algae such as \textit{Chlamydomonas reinhardtii} \citep{Omori2022} and bacteria such as \textit{E. coli} \citep{Hill2007}. 
Furthermore, investigations of the impact of a background flow are pertinent to the guidance of sperm cells in the female reproductive tract \citep{Kolle2009,Miki2013,Kantsler2014}, the design and control of microrobotic swimmers \citep{Nelson2010,Iacovacci2024}, and 
microbial contamination, infection, biofilm formation and ecology \citep{Mathijssen2019,Diluzio2005,Junot2019,Rusconi2015}. In turn, the prevalence and utility of background flows in microswimmer environments has motivated numerous theoretical studies investigating how background flows alter the swimmer dynamics. These range from studies closely aligned to observed microswimmer behaviours \citep{Kantsler2014,Ishimoto2015,Junot2019}
to more theoretical investigations that analyse the general dynamics and mechanics exhibited by theoretical models \citep{Hill2007,Zottl2012,Zottl2013,Chengala2013,Ishimoto2017,Ishimoto2023}. 

However, even among the theoretical studies, there has been a focus on specific background flows, especially Poiseuille and shear flows \citep{Zottl2012,Chengala2013,Ishimoto2015,Junot2019,Ishimoto2023}. Such flows are often well-motivated, since Poiseuille flows are common in a confined microgeometries and shear flows are a somewhat general approximation close to surfaces, noting that many swimmers accumulate near surfaces \citep{woolley2003,Lauga2006}. This focus on candidate flows prompts the broader question of how do microswimmers respond to more general background flows, especially in the absence of a confining geometry or a nearby surface. This is all the more relevant given individual models for swimmer dynamics are often and increasingly integrated into the development of models for collective behaviour, for example the works of  \citet{Saintillan2013,Ezhilan2013,Junot2019}. 

Consequently, a pertinent generalisation of previous studies is to document and classify the behaviour of microswimmers in more general flows. We pursue this in terms of the features of the flow and the swimmer's shape deformation cycle, also  termed a gait cycle below. However, the potential scope is unwieldy in full generality due to the diversity of possible flows and possible microswimmers. Thus, in this study, we restrict ourselves to planar, linear flows where, at most, we consider only spatially constant flows perpendicular to the plane of motion. The restriction to linear flows entails that the flow decomposes into a translation, rigid body rotation, and a pure strain, with the well-studied shear flow constituting an edge case in this general exploration. Accordingly, with mathematical precision required to balance the flow angular velocity and rate of strain to generate a pure shear flow, previous studies of shear flow may only represent edge cases for the possible dynamical behaviours of swimmers. 

Additionally, we restrict the range of possible swimmers to maintain tractability. One assumption is that the deformation cycle of the swimmer is sufficiently robust to be unchanged by the background flow. A further simplification that pertains to numerous microswimmers is high inefficiency: to swim, non-reciprocal body deformations are required for actuation, with the period of deformation giving one timescale, whilst the time to swim a body length gives a second timescale, with the latter being much longer in the case of swimmer inefficiency. This separation of timescales can be readily seen in biological swimmers, for instance sperm \citep{Smith2009a}, as well as many theoretical studies of idealised swimmers \citep{Curtis2013,Ishimoto2014a,Pak2015}. Hence, we assume such swimmer inefficiency, especially as it presents a means to extensively simplify the resulting equations of motion using the method of multiple timescales \citep{Bender1999}. 

We also assume that the velocity scale of the background flow is not extensively greater than the velocity scale of net swimming, so that the swimmer is not simply washed out. Nonetheless, we do relax this assumption for investigations of reciprocal swimming, where the swimmer oscillates back and forth with no net motion in a quiescent fluid, to consider whether or not the interaction of background flows with oscillatory swimmer motion can induce overall motility. We also allow for the prospect of oscillations of the background flow, noting there is an emerging interest in how swimmers interact with background flows \citep{Jo2016,Hope2016,Moreau2021b}, especially in scenarios where an inefficient swimmer makes progress in  a background flow oscillating with a frequency commensurate with that of the swimmer deformation \citep{Morita2018a,Morita2018b,Ishikawa2022}. The latter is particularly relevant to the current study and, thus, we incorporate background fluid flow oscillations with a frequency on the order of magnitude of the fast swimmer shape deformations. 

A more technical constraint is the restriction to swimmers with sufficient symmetry to ensure that the impact of the fluid rate of strain on the swimmer simplifies. In generality, this would be governed by two rank-three tensors \citep{Kim2005} and thus $2\times3^3=54$ degrees of freedom, each of which is a periodic function of the fast timescale since the swimmer is changing shape periodically to effect swimming. With symmetry constraints on the swimmer, including those required to ensure that simplifications from the planar symmetry of the flow are retained, these degrees of freedom can be reduced dramatically \citep{Ishimoto2020a}, which we document in more detail in \cref{geshs} below. For instance, swimmers that are bodies of revolution throughout their deformation cycle are special cases of the results considered below. However, we also explore the consequences of weaker symmetry constraints. For example, 
the presence of a swimmer-fixed axis with swimmer shape invariance to rotations of $2\pi/3$ about this axis, together with three invariant reflection planes containing the body fixed axis, throughout the swimmer shape deformation cycle is sufficient to apply the results of the analysis below, and other relatively low-symmetry shape deformation are considered too. 

Hence, our objectives are to analyse and classify the dynamics of inertialess but inefficient microswimming in linear background flows that are planar. In doing so, we retain sufficient generality to consider rotational, irrotational, and shear flows as special cases, with the general linear planar flow still amenable to analysis, even with flow oscillations on the same fast timescale as the swimmer's gait. From the perspective of rotational dynamics, we are particularly interested in whether a swimmer will tumble indefinitely, rock back and forth, or asymptote to a fixed angle, including how this is contingent on the properties of the both the swimmer and the flow, as well as interactions between them. Similar questions arise in considering translational dynamics, especially whether the swimmer inexorably drifts indefinitely across flow pathlines or settles into periodic orbits, and also whether swimmer-flow interactions can generate net motility for reciprocal swimmers and thus  break Purcell’s scallop theorem \citep{Purcell1977}. In turn, this allows us to consider to what extent, and how, a microswimmer may control its trajectory within a general planar linear flow. 

We pursue these objectives by first formulating the governing equations in \cref{Ges}. Then, utilising the assumption of inefficiency via its concomitant separation of timescales between swimmer undulation and motion, the governing equations are simplified in \cref{msa} and general features of the resulting solutions are examined. This is followed by numerical and theoretical investigations of special cases in \cref{sec: special cases}, with a focus on more symmetric swimmers and specific fluid flows for concreteness. Finally, we conclude with a general classification and summary in \cref{sec: discussion}.

\section{Governing equations}\label{Ges}
 
We derive the governing equations for inefficient swimmers undergoing rapid shape changing in planar background flows, with sufficient swimmer symmetry to render the swimmer motion planar and to ensure that interaction between the swimmer and the background flow rate of strain remains tractable. We additionally impose the following constraints:
the swimmer mechanics and background flow are effectively inertialess;
the swimmer shape is independent of the background flow; 
the swimmer shape oscillates on a fast timescale relative to the timescale of its net motion, with suitable generalisation if there is reciprocal swimming and, thus, no net motion; the background velocity field is a planar, incompressible Stokes flow that may in general oscillate on the fast timescale, commensurate with the timescale of the swimmer deformation oscillations. Whenever the background flow oscillates, for technical simplicity we additionally assume that the ratio of the flow oscillation period and the swimmer gait period is rational, with the overall period (the time for both the swimmer and the flow to return to the same phase) remaining a fast timescale. 

\begin{figure}
 \centering
 \includegraphics[width=0.65\textwidth]{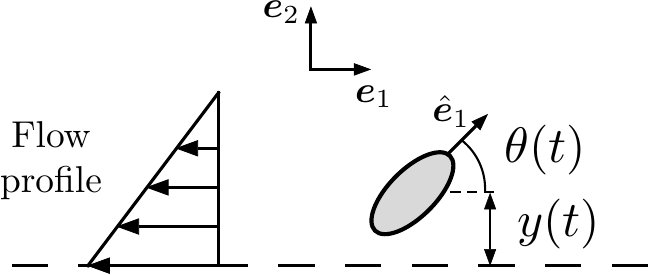}
 \caption{We illustrate a model swimmer in a planar, uni-directional, linear flow in the $x$-direction ($\e_1)$ and varying in the $y$-direction ($\e_2$), with the swimmer moving in the $xy$-plane. The swimmer orientation in the plane is captured via the unit vector $\ehat{1}$, which makes an angle $\theta$ with the $\e_1$ axis.}
 \label{fig1}
\end{figure}

With these assumptions, we immediately non-dimensionalise, with the viscosity scaled to unity by a choice of the pressure scale. In addition, we use the velocity, length and time-scales of the background flow, 
\begin{equation}\label{scles}  
    U_{bck}, \quad L_{bck}, \quad \tau_{b}=\frac{L_{bck}}{U_{bck}},
\end{equation} 
respectively, to remove the remaining dimensions, thus generating the non-dimensional framework for the governing equations that we work with below. The non-dimensional background flow is denoted by $\flowVel$. As an example, consider a non-dimensionalised planar linear shear flow of the form 
\begin{equation} \label{dimflow}
    \flowVel(\x,T) = \shearrate (T) y \e_1, 
\end{equation}
at a fixed instant in time, as depicted in \cref{fig1}, with $\{\e_1, \e_2, \e_3\}$ representing an inertial frame Cartesian basis for coordinates $x,y,z$ with $z$ out of the plane of the flow. Here, the non-dimensional shear rate, $\gamma(T)$, would be unity for shear that does not vary in time but, here, we allow it to oscillate on the fast timescale. This is represented by a dependence on a fast time variable $T$, associated with the swimmer gait oscillation and possible flow oscillation so that $T=2\pi$ corresponds to one fast period. In turn, $T$ is related to the slow time variable, $t$, via 
\begin{equation}\label{Twt} 
    T=\omega t, \quad \omega \gg 1.
\end{equation} 
We also require $t$ to be commensurate with the net swimming timescale, so that $\tau_{b}$ is of the order of the time it takes for the swimmer to have a net translation of a body length, so that the swimmer is not washed out by the flow. This also immediately satisfies our assumption that the swimmer is inefficient given $\omega \gg 1$, and can be viewed as the definition of inefficiency. However, a minor refinement is needed when considering the timescales for reciprocal swimmers, which never translate a body length in a quiescent fluid. In particular, reciprocal swimmers are, in a suitable sense, maximally inefficient but the net swimming timescale is nevertheless ill-defined. Thus, above, we have not used the time to swim one body length for non-dimensionalisation and, in addition, for reciprocal swimmers we relax the  requirement that $\tau_{b}$ is on the timescale needed for the swimmer to translate a body length.

\subsection{Flow kinematics and general governing equations}
\label{fkge}

With the timescales defined, we can proceed to consider the background flow in detail and derive the governing equations for the rotational and translation dynamics of the swimmer. We have $ \x = x \e_1+y\e_2+z\e_3$ as the non-dimensional position of a general point
with respect to the laboratory-fixed basis, $\{\e_1, \e_2, \e_3\}$; analogously, the swimmer-fixed frame has a basis given by $\{\ehat{1}, \ehat{2}, \ehat{3}\}$ with its origin $\x_c$ at the centroid of the swimmer. As further detailed below, the swimmer is taken to possess a body-fixed symmetry axis throughout its deformation cycle, which we take to be aligned in the body-fixed direction $\ehat{1}$. Noting the assumptions of flow planarity and sufficient swimmer symmetry to ensure planar motility (detailed below) we can, with suitable initial conditions implicitly assumed, take the swimmer axis of symmetry to lie in the plane of the flow which, without loss, is the $xy$-plane of the inertial frame. Note that this entails that $\ehat{1}$ also lies in this plane. Hence, we have the simple relations between the basis vectors 
\begin{equation} \label{ct} \ehat{1}=\cos\theta \e_1 +\sin\theta \e_2,
 \quad\ehat{2}=-\sin\theta \e_1 +\cos\theta \e_2,\quad\ehat{3}=\e_3,
\end{equation} with $\theta$ as depicted in Figure \eqref{fig1}. In practice, this alignment also requires stability of planar swimming, which we do not explore in this work. 

Neglecting the influence of the swimmer on the flow (so that no-slip conditions on the swimmer surface are not imposed), we denote the linear, planar, non-dimensional background flow field as $\flowVel(\x,T)$ with 
rate of strain and angular velocity given by 
\begin{equation} \label{rosavnew}
    \flowStrainRate (T) = \frac{1}{2} \left[\vnabla \flowVel + (\vnabla \flowVel)\transpose\right] , \quad
\flowAngVel (T) = \frac{1}{2} \vnabla \wedge \flowVel, 
\end{equation} 
respectively, where $\cdot\transpose$ denotes the transpose. As above, the prospect of rapid background flow oscillation is indicated by the fast time variable dependence of the rate of strain and angular velocity while, by the linearity of the flow, we have $\flowAngVel$ and $\flowStrainRate$ are independent of spatial location. As the flow is planar we have the further simplifications 
\begin{equation}
    \flowAngVel (T) = \flowAngSpeed(T)\ehat{3} = \flowAngSpeed(T) \e_3, \quad \flowAngSpeed (T) = \frac{1}{2} \e_3\cdot \vnabla \wedge \flowVel .
\end{equation}
Further, we have that the rate of strain tensor, $ \flowStrainRate$, can be written with respect to the laboratory basis as
\begin{equation}\label{Ecomp}
    \flowStrainRate = \begin{bmatrix}
    E^{11}(T) & E^{12}(T) & 0\\
    E^{12}(T) & -E^{11}(T) & 0\\ 
    0 & 0 & 0
 \end{bmatrix},
\end{equation}
noting that symmetry and flow incompressibility entail that only two degrees of freedom remain. In the same laboratory reference frame, the background flow takes the explicit form 
\begin{align}
 \flowVel(\x,T) &= \flowVel_{tr}(T)+\underbrace{(-y\e_1 + x\e_2)\flowAngSpeed(T)}_{\text{pure rotation}} \nonumber\\
 &+ \underbrace{(x\e_1 -y\e_2)E^{11}(T)  + (y\e_1 + x\e_2)E^{12}(T)}_{\text{pure strain}} \nonumber\\ 
 &= \flowVel_{tr}(T)+ 
\flowAngVel(T)\wedge \x + \flowStrainRate(T)\x , \label{ubcklin}
\end{align}
where $\flowVel_{tr}(T)$ is a translational flow that has no spatial dependence. It is useful to note that velocity field satisfies the identity 
\begin{eqnarray}\label{exp0}
\flowVel(\x,T) = \flowVel_c + \flowAngVel(T)\wedge (\x-\x_c) + \flowStrainRate(T)(\x-\x_c)
\end{eqnarray} 
by linearity, where $\flowVel_c = \flowVel(\x_c,T)$ is the background flow at the centroid $\x_c$ of the swimmer.

With the background flow specified, we show in \cref{appendix1} that the planar motion of the swimmer is governed by 
\begin{align} \label{xeqn0} 
 \frac{\mathrm{d}\x_c}{\mathrm{d}t} &= \flowVel_c + \omega U(T) \ehat{1}+ P\ehat{1} -\tilde{\tensor{g}}\flowStrainRate , \\ 
 \label{teqn0} \dot \theta \e_3 =\dot \theta \ehat{3} 
 &= \Omega^*(T) \ehat{3}+  \omega\Omega_f(T)\ehat{3}+ \Omega(T)\ehat{3}- \tilde{\tensor{h}} \flowStrainRate.
\end{align}
Here, $\omega U(T)$, $P$, and $\omega\Omega_f(T) + \Omega(T)$ are the oscillatory swimming speed, average progressive swimming speed, and rate of rotation of the swimmer in the absence of any background flow, each of which are assumed to be given. Note that we have decomposed the latter into a fast, $\bigO{\omega}$ component $\Omega_f(T)$ and an $\bigO{1}$ component $\Omega(T)$ for later convenience. As described in detail in \cref{appendix1}, we assume $P\sim\bigO{1}$ and $\omega U(T)\sim\bigO{\omega}$, with the average of $U(T)$ evaluating to zero over a period of the fast oscillation, $T= 2\pi$. Note that setting $\flowVel_c$, $\flowStrainRate$ and $\Omega^*$ to zero gives the equations of motion in the absence of flow. The terms $\tilde{\tensor{g}}$ and $\tilde{\tensor{h}}$ are rank three tensors that capture how the rate of strain of the background influences the swimmer dynamics. The assumption of planarity requires $\tilde{\tensor{h}} \flowStrainRate \parallel \ehat{3}$, which we impose by constraining the swimmer shape throughout its gait cycle to have sufficient symmetry. However, we remark that this framework retains validity even when $\tilde{\tensor{g}}\flowStrainRate$ has a component in the $\ehat{3}=\e_{3}$ direction, so we do not disallow this prospect a priori. Here and throughout, all angular velocities are constrained to the $\ehat{3}=\e_{3}$ direction by the assumption of planarity. 

Below, we simplify these equations of motion by enforcing geometrical symmetries of the swimmer. The generality of the resulting derivations and equations naturally requires a relatively large number of parameters and variables. Hence, we summarise these in \cref{Tab1,Tab1.1} for reference, together with the parameters and variables already introduced.

\subsection{Simplified governing equations} \label{geshs}
In order to be consistent with our assumption of planarity, we must restrict ourselves to particular classes of swimmer geometry. In full generality, this is necessarily technical and requires significant notation. A reader seeking a concrete example might consider the swimmer to be a body of revolution with fore-aft symmetry, though we remark that much more general geometries are admissible within the present framework. Below, we elaborate on the details of some additional cases, though these can be safely skipped if one is willing to accept the presence of the time-dependent geometrical parameters $B(T)$, $\lambda_5(T)$, $\eta_2(T)$, $\eta_3(T)$, and $\eta_4(T)$ in the explicit and simplified governing equations:
\begin{multline}\label{xeqnf}
    \diff{\x_c}{t} =  \flowVel_c + \omega U(T) \ehat{1} + P\ehat{1}  - \eta_2 (T)[\hat{\e}\transpose_1 \flowStrainRate (T)\hat{\e}_2 ]   \hat{\e}_3 \\  +\eta_3(T)\vec{B}_0(T,\sin2\theta,\cos2\theta)\hat{\e}_1
    -\eta_4 (T)[\hat{\e}\transpose_2 \flowStrainRate (T)\hat{\e}_2 ] \hat{\e}_2,
\end{multline}
\begin{multline}\label{eq: intermediate orientation1}
    \diff{\theta}{t} = \flowAngSpeed(T)+   \omega \Omega_f(T)+\Omega(T)  
    +[\lambda_5(T)E^{12}(T)-B(T)E^{11}(T)]\sin2\theta \\
    +[\lambda_5(T)E^{11}(T)+B(T)E^{12}(T)]\cos2\theta.
\end{multline}
Here, $B(T)$ is the Bretherton parameter and, using $c\equiv \cos2\theta$ and $s\equiv \sin2\theta$, we have that  $\vec{B}_0(T,\sin2\theta,\cos2\theta)$ is given by 
\begin{multline}\label{B0def}
    \vec{B}_0(T,\sin2\theta,\cos2\theta)
    = \flowStrainRate- [E^{11}(T)\cos2\theta +E^{12}(T)\sin2\theta](\e_1\e_1\transpose +\e_2\e_2\transpose )
    \\  = 
    \begin{pmatrix}
    E^{11}(T)(1-c) - E^{12}(T)s & E^{12}(T) & 0 \\
    E^{12}(T) & -[E^{11}(T)(1+c) + E^{12}(T)s] & 0 \\ 0 & 0 & 0 
    \end{pmatrix},
\end{multline}
where the final expression is  with respect to the laboratory basis. In the remainder of this section, we explicitly describe the construction of these governing equations from \cref{xeqn0}; we continue with an analysis of these equations in \cref{msa}.

 \setlength\extrarowheight{3pt}
\begin{table}
\smaller
\begin{NiceTabular}{|C{2.5cm}|m{10.5cm}|}
\hline 
 Parameter/Variable & Description\\
 \hline\hline 
$U_{bck}$, $L_{bck}$, $ \tau_{b}=L_{bck}/U_{bck}$ & Dimensional velocity, length, and timescale of the background flow.  \cref{scles}. \\ \hline
 $t$, $T=\omega t $ & Slow and fast timescales, respectively, with $\omega \gg 1$. See \cref{Twt}. \\ \hline
 $\{\e_1, \e_2, \e_3\}$ & Laboratory-fixed basis. \cref{fkge}.\\ \hline
$\{\ehat{1}, \ehat{2}, \ehat{3}\}$ & Swimmer-fixed basis. Note $\ehat{3}=\e_3$. See \cref{fkge}. \\ \hline
$\theta, \theta_0$ & Swimmer orientation angle. See \cref{ct}. \\ \hline
$\x$, $\x_c$ & Field point and the swimmer centroid. See \cref{fkge}.\\ \hline
$\flowVel(T)$, $\flowStrainRate(T)$, $\vec{\Omega}^*(T)=\Omega^* \e_3$ & Background flow, its rate of strain tensor and angular velocity. See \cref{decomp,rosavnew}.\\ \hline
 $\vec{\Omega}(T,\vec{0})= \omega\Omega_f(T) \e_3 + \Omega(T) \e_3 $ & Angular velocity of the swimmer when $\flowVel=\vec{0}$. See \cref{decomp,rosavnew}.\\ \hline
$E^{ij}$, $\hat{E}^*_{ij}$ & Components of $\flowStrainRate$ in the laboratory and swimmer frames. See \cref{Ecomp,hatE}.  \\ \hline
$\flowVel_c$, $\flowVel_{tr}$ & Background flow at the swimmer centroid and its spatially constant contribution. See \cref{exp0,ubcklin}. \\ \hline
$P$, $U(T)$, $U_I(T)$ & Mean swimming speed, oscillatory swimming speed and its integral. See \cref{xeqn0,svage,veleq}.  \\ \hline
$C_{nv}$, $C_{nh}$, $D_n$, $D_{nh}$ & Types of helicoidal symmetry. See \cref{geshs}.\\ \hline
$\tilde{\tensor{g}}$, $\tilde{\tensor{h}}$ & Rank 3 tensors capturing the impact of the rate of strain on motility. See  \cref{gmt1}. \\ \hline
$\vec{d}_1, \vec{d}_2, \vec{d}_3, \vec{d}_4, \vec{d}_5$ & Vectors used to decompose $-\tilde{\tensor{g}}\flowStrainRate$, $-\tilde{\tensor{h}}\flowStrainRate$. See \cref{decomp,geshs}. \\  \hline
$\lambda_2$, $\lambda_5$, $\eta_2$, $\eta_3$, $\eta_4$ & Coefficients of the above decomposition for $-\tilde{\tensor{g}}\flowStrainRate$ and  $-\tilde{\tensor{h}}\flowStrainRate$. See \cref{decomp,geshs}. \\  \hline
$B=-\lambda_2$ & Bretherton shape parameter. See \cref{decomp,geshs}.  \\  \hline
$\vec{B}_0(T,\theta) $ & Matrix used to summarise translational equation of motion before multiple scales approximation. See \cref{B0def}.  \\ \hline 
\end{NiceTabular}
\caption{A list of parameters and variables used in the formulation of the governing equations, including the description of the background flow and the swimmer symmetries. All are non-dimensional except for the first row of scales used to non-dimensionalise the system. Note that the variable  $\flowVel_{tr}$ is overloaded and relative to either the 2D flow plane or 3D more generally according to context, with the 3D expression including the constant $z$-contribution to the background flow. Parameters and variables introduced in the Appendices that do not appear in the main text are not listed.\label{Tab1}}
\end{table}

\setlength\extrarowheight{3pt}
\begin{table}
\smaller
\begin{NiceTabular}{|C{4.0cm}|m{9.0cm}|}
\hline  
		Parameter/Variable & Description\\
		\hline\hline
		 $\x_0$, $\theta_0$, $\x_1$, $\theta_1,$  & Leading order, and next to leading order,  approximation to the swimmer centroid, orientation angle. See \cref{pser}.\\   \hline
		$a_*$, $b_*$, $b_{**}$, $c_*$, $c_{**}$, $a$, $b$, $c$  & Terms summarising contributions to the angular equation of motion. $a$, $b$, $c$ are fast-timescale averages of $a_*$, $b_{**}$, $c_{**}$. See \cref{abc,dthetadt,cstarstar}. \\   \hline
		$p$, $q$, $\theta_{00}$ & $p=(b^2+c^2-a^2)^{1/2}$, $q=(a^2-b^2-c^2)^{1/2}$, $\theta_{00}=\theta(t=0)$. See \cref{tsol}. \\ \hline
		$\ehat{10}(\theta_0)$, $\ehat{20}(\theta_0)$, $\ehat{10}(\avg{\theta}_0)$, $\ehat{20}(\avg{\theta}_0)$ & Leading-order multiple scales approximation to $\ehat{1}$, $\ehat{2}$, as functions of either $\theta_0$ or $\avg{\theta}_0$. See \cref{ehat0,ehat00}.\\ \hline
		$\Omega_{fI}(T)$, $\Psi(T)$ & The fast time-scale integral of $\Omega_f(T)$, $\Psi(T)=\Omega_{fI}(T)-\avg{\Omega}_{fI}$. See \cref{OmfI,Psi}. \\ \hline
		$U_{cI}(T)$, $U_{sI}(T)$ & The fast time-scale integrals of $U(T)\cos\Psi(T),$ ~$U(T)\sin\Psi(T)$. See \cref{UcI}. \\ \hline
		$\Phi_c(T),$ $\Phi_s(T)$& $U_{cI}(T)-\avg{U}_{cI},~U_{sI}(T)-\avg{U}_{sI}$. See \cref{Phic}.\\ \hline
		$\mat{\Lambda}(T)$ & Matrix used in summarising expansion of background flow. \cref{Lambda}. \\ \hline
		$\mathcal{L}, \mathcal{G} $ & Linear operator and image for the Fredholm alternative. See \cref{fat}. \\ \hline
		$\vec{p}_1,\vec{p}_2,\vec{p}_3, \vec{p}_4$ &Spanning basis for the nullspace of the adjoint of $\mathcal{L}$. See \cref{nsbasis}.\\ \hline
		$\chi$ & $\chi =a_*(T)-b_{**}(T)\sin(2\avg{\theta}_0)-c_{**}\cos(2\avg{\theta}_0)$. See \cref{chi}.\\ \hline
		$M_1,M_2,M( \sin2 \avg{\theta}_{0} ,\cos2 \avg{\theta}_{0} )$  & $M$ is a linear  function, with constants $M_1,M_2$ summarising $\ehat{10}(\avg{\theta}_0)$ dependence of translational solution. See \cref{eqnxc00}. \\ \hline
		$N_1,N_2,N( \sin2 \avg{\theta}_{0} ,\cos2 \avg{\theta}_{0} )$  & $N$ is a linear  function, with constants $N_1,N_2$ summarising $\ehat{20}(\avg{\theta}_0)$ dependence of translational solution. See \cref{eqnxc00}. \\ \hline
		$\vec{C}( \sin2 \avg{\theta}_{0} ,\cos2 \avg{\theta}_{0}, \sin \avg{\theta}_{0} ,\cos \avg{\theta}_{0})$  & $\avg{\theta}_0$-dependent vector within the translational equations. See \cref{solxc00}. \\ \hline
		$\mat{A}$, $\mat{K} = \exp[\mat{A}t]$ & Constant matrix and its exponential for the translational equation of motion and its solution. See \cref{eqnxc00,solxc00}.\\ \hline
		$\nu$, $\mu$ & $\nu = (\avg{E^{11}}^2+\avg{E^{12}}^2-\avg{\Omega^*}^2)^{1/2}$, $\mu = (\avg{\Omega^*}^2-\avg{E^{11}}^2-\avg{E^{12}}^2)^{1/2}$ describing the in-plane dynamics of the translational motion. See \cref{solxc00}. \\ \hline
		$\mathcal{A}$, $\mathcal{B}$, $\mathcal{K}$ & Constants in the trajectory equation for an oscillatory shear background flow. See \cref{Acal,Kcal}. \\ \hline 
	\end{NiceTabular}
	\caption{A list of parameters and variables describing the multiscale simplifications and aspects of the explicit solutions to the governing equations for special cases. They are all non-dimensional. An overline of any variable refers to taking a temporal average over a period of the fast timescale, as defined by \cref{tempavg}. Note that the variables $\bar{\x}_0$ and $\mat{\Lambda}$ are overloaded and relative to either the 2D flow plane, or 3D more generally, according to context, with 3D expressions respectively including the $z$-contribution to the leading order swimmer centroid position and a trivial zero-padding in the third dimension when $\mat{\Lambda}$ acts on a three-dimensional vector. Parameters and variables introduced in the Appendices that do not appear in the main text are not listed in this table. \label{Tab1.1}}
\end{table}

\subsection{Simplification of the governing equations and detailed geometrical constraints}\label{sgedgc}
In order to ensure that the swimmer only rotates in the plane of the flow, we immediately restrict consideration to swimmers whose shape throughout the gait cycle possesses a rotational symmetry of degree $n\geq 3$. That is, the swimmers possess a body fixed axis throughout the gait cycle such that there is a shape invariance to rotations around this axis of angle $2\pi/n$ \citep{Ishimoto2020a}. 
In Shoenflies notation, such a body with $n$-fold rotational symmetry is denoted by $C_n$. In turn, the body symmetry enforces the constraints on the entries of the third-rank tensors, $\tilde{\tensor{g}}$ and $\tilde{\tensor{h}}$, yielding another type of shape classification based on the symmetry of these tensors.
This symmetry is a \emph{hydrodynamic symmetry} and we refer the interested reader to the detailed definitions of \citet{Ishimoto2020b, Ishimoto2020a, Ishimoto2023} for further elaboration on this rich topic. With this $C_n$ ($n\geq 3$) body symmetry, \citet{Ishimoto2020a} considers the structure of $-\tilde{\tensor{g}} \flowStrainRate$ and $-\tilde{\tensor{h}} \flowStrainRate$ and shows that a $C_n$ ($n\geq 3$) body has {\it helicoidal symmetry of degree 3}, for which the body dynamics are explicitly written down in the same form. By taking $n\rightarrow\infty$, it is known that a simple helix approximately follows the same dynamical equations \citep{Ishimoto2020b}. However, the helicoidal symmetry alone is \emph{not} sufficient for our purposes. Thus, we closely follow \citep{Ishimoto2020a} to determine $-\tilde{\tensor{g}} \flowStrainRate$ and $-\tilde{\tensor{h}} \flowStrainRate$, noting that additional simplifications will arise here from both the planar nature of the flow and the restriction of swimmer shapes to those for which $\tilde{\tensor{h}} \flowStrainRate \parallel \ehat{3}$. Before doing so, it is convenient to introduce the notation 
\begin{equation}\label{hatE} 
 \hat{E}^*_{ij} (T,\sin2\theta,\cos2\theta)\coloneqq \hat{\e}\transpose_i \flowStrainRate (T)\hat{\e}_j = \hat{E}^*_{ji}(T,\sin2\theta,\cos2\theta)
\end{equation}
for $i,j\in\{1,2,3\}$. Explicitly, for $j\in\{1,2,3\}$ we have 
\begin{subequations}\label{Estar}
\begin{align}
 \hat{E}^*_{11}&= - \hat{E}^*_{22} = E^{11}(T) \cos2\theta +E^{12}(T)\sin2\theta, \\
 \hat{E}^*_{12} &= E^{12} (T)\cos2\theta -E^{11}(T)\sin2\theta,\\
 \hat{E}^*_{3j}&=0,
 \end{align}
 \end{subequations}
 each functions of $T$, $\sin2\theta$, and $\cos2\theta$.
 
With our assumptions and simplifications, we can now decompose $-\tilde{\tensor{g}} \flowStrainRate$ and $-\tilde{\tensor{h}} \flowStrainRate$ via
\begin{subequations}\label{decomp}
\begin{align} 
 -\tilde{\tensor{g}} \flowStrainRate &= \eta_2(T)\vec{d}_2(T,\theta)+\eta_3(T)\vec{d}_3(T,\theta)+\eta_4(T)\vec{d}_4(T,\theta), \\
 -\tilde{\tensor{h}} \flowStrainRate &= \lambda_2(T)\vec{d}_2(T,\theta)+\lambda_5(T)\vec{d}_5(T,\theta)
\end{align}
\end{subequations}
with
\begin{subequations}
\begin{align}
\vec{d}_2(T,\theta)= -\hat{E}_{12}^*\ehat{3}, \quad \vec{d}_3(T,\theta) = (\flowStrainRate-\hat{E}^*_{11} \mat{I})\ehat{1} , \\ \vec{d}_4(T,\theta)= -\hat{E}_{22}^*\ehat{2}, \quad \vec{d}_5(T,\theta) = -\hat{E}_{22}^*\ehat{3},
\end{align}
\end{subequations}
where 
$\lambda_2(T)\equiv -B(T)$, $\lambda_5(T)$, $\eta_2(T)$, $\eta_3(T)$, and $\eta_4(T)$ are shape dependent parameters. 

Notably, symmetries have to be imposed on the swimmer in order to ensure that 
there are no contributions to $ -\tilde{\tensor{h}} \flowStrainRate$ from 
$\vec{d}_3(T,\theta)$ and $\vec{d}_4(T,\theta)$, so that $\tilde{\tensor{h}} \flowStrainRate \parallel \ehat{3}$. For the swimmer shapes we consider throughout, this also ensures that $ -\tilde{\tensor{g}} \flowStrainRate$ has no contribution from $\vec{d}_5(T,\theta)$ \citep{Ishimoto2020a}. 
Thus, we do not consider the full range of shapes for which \cref{decomp} is valid, but instead require additional restrictions. Particular examples of swimmer shapes with sufficient symmetry to be admissible within the present framework, together with the related restrictions on \cref{decomp}, are presented in detail in \cref{app: shapes} for the interested reader. Finally, we note that for the simple canonical example of a body of revolution with fore-aft symmetry, there is extensive simplification, with $\lambda_5=\eta_2=\eta_3=\eta_4=0.$ Such a swimmer does not have the asymmetry needed to generate rotation in the absence of a flow, so that its angular velocity in the absence of flow, $\omega \Omega_f(T)+\Omega(T),$ is also zero.

\section{Multiscale analysis in time-dependent flows}\label{msa}
We proceed to use a multiscale analysis to simplify the governing equations, taking advantage of the separation of timescale arising from $\omega\gg 1,$ so that $\omega t$ is fast timescale relative to $t$.   As previously noted, we implicitly assume that the period of the fast timescale oscillations is a small integer number of periods of any background flow oscillation and swimmer deformation oscillation, including treadmilling, which ensures there is only one fast timescale.  

\subsection{Multiple scales for the angular dynamics}
Defining
\begin{subequations}\label{abc}
\begin{align}
a_*(T)&=\flowAngSpeed(T)+\Omega(T),\\
b_*(T)&=B(T)E^{11}(T)-\lambda_5(T)E^{12}(T),\\
c_*(T)&=-B(T)E^{12}(T)-\lambda_5(T)E^{11}(T), 
\end{align}
\end{subequations}
for notational convenience, the angular evolution equation becomes
\begin{equation}\label{dthetadt}
 \diff{\theta}{t} = \omega \Omega_f(T) + a_*(T) - b_*(T)\sin{2\theta} -c_*(T)\cos{2\theta} , 
\end{equation}
which is decoupled from the equations for translational motion and thus may be treated in isolation.

To study this angular dynamics, we use the method of multiple timescales, exploiting $\omega \gg 1$. The slow timescale, $t$, is associated with the flow, and the fast timescale, $T=\omega t$, is associated with the swimmer deformation and treadmilling. Hence, the total time derivative decomposes via
\begin{equation}
\diff{}{t} = \pdiff{}{t} + \omega 
\pdiff{}{T}.
\end{equation}
With a zero subscript denoting the leading order, we expand $\theta$ via 
\begin{equation}
\theta = \theta_0(t,T) + \frac{1}{\omega} \theta_1(t,T) + \ldots,
\end{equation}
with $\theta_1(t,T)$ inheriting the $2\pi$-periodicity of the fast time dynamics, as is standard in the multiple timescales method. Thus, at $\bigO{\omega}$ and $\bigO{1}$ we have, respectively, 
\begin{subequations} \label{t1eq} 
\begin{align}
 \theta_{0T} &= \Omega_f(T),\\
 \theta_{1T} &= -\theta_{0t}+ a_*(T) - b_*(T)\sin{2\theta_0} -c_*(T)\cos{2\theta_0}.
\end{align}
\end{subequations}
This gives 
\begin{equation}\label{OmfI}
 \theta_0 (t,T)= \Omega_{fI}(T) + \theta_{00}(t), \quad \Omega_{fI}(T) = \int_0^T \Omega_{f}(S) \intd{S},
\end{equation} 
where $\theta_{00}(t)$ is an undetermined function of $t$ alone. For convenience, we denote the fast-timescale-average of a quantity via a bar, that is 
 \begin{equation}\label{tempavg} 
 \avg{Q} \coloneqq \frac 1 {2\pi} \int_0^{2\pi} Q(t,T) \intd{T},
 \end{equation} 
so that averaging \cref{OmfI} gives
\begin{equation}
 \avg{\theta}_0 (t)= \avg{\Omega}_{fI} +  \theta_{00}(t).
\end{equation}
Eliminating $\theta_{00}(t)$ gives
\begin{equation}\label{Psi}
\theta_0 (t,T) = \avg{\theta}_0 (t) + \Psi(T), \quad \Psi(T) = 
\Omega_{fI}(T) - \avg{\Omega}_{fI},
\end{equation}
from which we have
$\avg{\Psi}=0$ and $\theta_{0t} = \mathrm{d}\avg{\theta}_0/\mathrm{d}t$.
Hence, we can expand
\begin{align}  
b_*(T)\sin{2\theta_0} +c_*(T)\cos{2\theta_0} &= b_*(T)\sin(2\avg{\theta}_0 + 2\Psi) +c_*(T)\cos(2\avg{\theta}_0 + 2\Psi) \nonumber\\ 
&=  b_{**}(T)\sin(2\avg{\theta}_0) +c_{**}(T)\cos(2\avg{\theta}_0), \label{bstarstar}
\end{align}
 with, on recalling the definitions of \cref{Estar}, 
\begin{subequations}\label{cstarstar}
\begin{align}
 b_{**}(T) &= B(T)\hat{E}^*_{11}(T,\sin(2\Psi),\cos(2\Psi))-\lambda_5(T)\hat{E}^*_{12}(T,\sin(2\Psi),\cos(2\Psi)) \\ 
c_{**}(T) &= -B(T)\hat{E}^*_{12}(T,\sin(2\Psi),\cos(2\Psi))-\lambda_5(T)\hat{E}^*_{11}(T,\sin(2\Psi),\cos(2\Psi)).
\end{align} 
\end{subequations}
Hence, the impact of the fast angular dynamics is that the rate of strain tensor contributions are taken only  after a rotating the basis by an angle $\Psi(T)$. 

As the only homogeneous solution of the equation for $\theta_1$ (given by \cref{t1eq}) that also satisfies the requirement of periodicity is the constant solution, \cref{bstarstar} and the Fredholm alternative theorem give 
\begin{equation} \label{fat0} \int_0^{2\pi} \left\{\theta_{0t}+a_*(T) - b_{**}(T)\sin{2\avg{\theta}_0} -c_{**}(T)\cos{2\avg{\theta}_0} \right\} \intd{T}=0.
\end{equation} 
Thus, the leading order dynamics is governed by the simpler differential equation 
\begin{equation} \label{teqnf} \dfrac{\mathrm{d} \avg{\theta}_0}{\mathrm{d} t} = a - b \sin 2\avg{\theta}_0 - c \cos2\avg{\theta}_0 \, , 
\quad a=\avg{a}_*, \, b=\avg{b}_{**} , \, c= \avg{c}_{**}. 
\end{equation} 
Note that we have defined $a,b,c$ as the averages of $a_*,b_{**},c_{**}$, respectively, for ease of notation in what follows. These quantities will be key in exploring the emergent behaviour in \cref{sec: features: rotation,sec: features: translation,sec: special cases}.
With the initial condition that 
\begin{equation}        \theta_0(t=0)=\avg{\theta}_0(t=0)=\theta_{00}
\end{equation}
and the definitions 
\begin{equation}\label{tsolpq} 
 p=(b^2+c^2-a^2)^{1/2}, \quad q=(a^2-b^2-c^2)^{1/2},
\end{equation}
one can readily determine 
\begin{equation} \label{tsol} 
\tan \avg{\theta}_0 = 
\left\{
\begin{array}{lr}
\frac{p\tan\theta_{00}+[a-c-b\tan\theta_{00}]\tanh(pt)}{p+[b-(a+c)\tan\theta_{00}]\tanh(pt)} , &  a^2 < b^2 + c^2, \\  & \\ 
\frac{\tan\theta_{00} + [a-c-b\tan\theta_{00}]t}{1+[b-(a+c)\tan\theta_{00}]t} , & a =(b^2 + c^2)^{1/2} , \\ & \\
\frac{q\tan\theta_{00} + [a-c-b\tan\theta_{00}]\tan(qt)}{q+[b-(a+c)\tan\theta_{00}]\tan(qt)} ,  & a^2 > b^2 + c^2 ,
\end{array} 
\right.
\end{equation} 
where the appropriate branches of $\arctan$ are chosen so that $\avg{\theta}_0$ is continuous. 

\subsection{Multiple scales for the translational dynamics}
We proceed to consider the translational dynamics by applying the multiple scales method to \cref{xeqnf,xeqnf1} using the expansions 
\begin{subequations}\label{pser}
\begin{align} 
    \x_c &= \x_{0}+\frac 1 \omega \x_{1} +\ldots,\\
    \theta &= \theta_0(t,T) + \frac 1 \omega \theta_1(t,T) + \ldots = \avg{\theta}_0(t) + \Psi(T)+ \frac 1 \omega \theta_1(t,T) + \ldots. 
\end{align}
\end{subequations}
With $\cdot\transpose$ denoting the transpose, we also define
\begin{equation}\label{ehat0}
\ehat{10}(\avg{\theta}_0)\coloneqq
[\cos\avg{\theta}_0(t),\sin\avg{\theta}_0(t)]\transpose,  \quad
\ehat{20}(\avg{\theta}_0)\coloneqq[-\sin\avg{\theta}_0(t),\cos\avg{\theta}_0(t)]\transpose,
\end{equation}
so that 
\begin{subequations}\label{ehat00}
\begin{align}
    \ehat{10}(\theta_0) &\coloneqq[\cos\theta_0,\sin\theta_0]\transpose = \cos\Psi \ehat{10}(\avg{\theta}_0)+\sin\Psi\ehat{20}(\avg{\theta}_0) \\ 
    \quad\ehat{20}(\theta_0)&\coloneqq [-\sin\theta_0,\cos\theta_0]\transpose =\cos\Psi\ehat{20}(\avg{\theta}_0) -\sin\Psi \ehat{10}(\avg{\theta}_0).
\end{align}
\end{subequations}
We also have immediately have
\begin{equation}
    \diff{\ehat{10}}{t}(\avg{\theta}_0) = \diff{\avg{\theta}_0}{t}\ehat{20}(\avg{\theta}_0), \quad \diff{\ehat{20}}{t}(\avg{\theta}_0) =-
 \diff{\avg{\theta}_0}{t}\ehat{10}(\avg{\theta}_0).
\end{equation}

Then, at the leading order of the multiple scales expansion for the translational governing equations, we have
\begin{equation}
    \pdiff{\x_0}{T} = \gait(T)\ehat{10}(\theta_0),
\end{equation}
so that $\x_0$ and its average are given by
\begin{subequations}\label{xc00}
\begin{align}  
 \x_{0} (t,T) &= \x_{00}(t)+U_{cI}(T) \ehat{10}(\avg{\theta}_0)+U_{sI}(T) \ehat{20}(\avg{\theta}_0), \\
 \avg{\x}_{0} (t) &= \x_{00}(t)+\avg{U}_{cI} \ehat{10}(\avg{\theta}_0)+\avg{U}_{sI} \ehat{20}(\avg{\theta}_0),
\end{align}
\end{subequations}
 where $\x_{00}(t)$ is a to-be-eliminated function of integration and
 \begin{equation}\label{UcI}
 U_{cI}(T)=\int_0^T U(S)\cos(\Psi(S))\intd{S}, \quad U_{sI}(T)=\int_0^T U(S)\sin(\Psi(S))\intd{S}.
 \end{equation}
For future use below, we note the assumption of no net swimming on the fast timescale entails that we have $\avg{U}_{cI},\avg{U}_{sI}=\bigO{1/\omega}$. In other words, the interaction of fast velocities and fast turning still average to zero for $\x_{0} (t,T)$ on the fast timescale in \cref{xc00}, rather than induce a net velocity on the fast scale that breaks the assumption of swimmer inefficiency.

We continue by writing $\x_0(t,T)$ as the sum of its average and oscillatory, zero-mean terms via
\begin{equation}\label{xxavgfast} 
    \x_{0}(t,T)=\avg{\x}_0(t)+ \Phi_c(T) \ehat{10}(\avg{\theta}_0) + \Phi_s(T) \ehat{20}(\avg{\theta}_0),
\end{equation}
where 
\begin{equation}\label{Phic} 
    \Phi_c(T) = U_{cI}(T)-\avg{U}_{cI}, \quad \Phi_s(T) = U_{sI}(T)-\avg{U}_{sI}
\end{equation}
and, notably, $\avg{\Phi}_c=\avg{\Phi}_s=0$. In practice the $\avg{U}_{cI}$, $\avg{U}_{sI}=\bigO{1/\omega}$ terms can be safely dropped.  

Given that the background flow is an affine map of the position vector from \cref{ubcklin}, we have at leading order, 
\begin{multline} 
    \flowVel_c = \flowVel( \x_{0} (t,T),T ) = \flowVel_{tr}(T) +\mat{\Lambda}(T) \avg{\x}_{0}(t)+ \Phi_c(T)\mat{\Lambda}(T)\ehat{10}(\avg{\theta}_0) \\ 
    + \Phi_s(T)\mat{\Lambda}(T)\ehat{20}(\avg{\theta}_0), 
\end{multline}
with
\begin{equation}\label{Lambda}
    \mat{\Lambda}(T) \coloneqq 
    \begin{bmatrix}
        E^{11}(T) & E^{12}(T) -\Omega^*(T)& 0\\
        E^{12}(T) +\Omega^*(T)& -E^{11}(T) & 0\\ 
        0 & 0 & 0
    \end{bmatrix}
\end{equation}
written with respect to the laboratory basis. Hence, at the next order we have
\begin{equation}
\begin{split}
    \pdiff{\x_1}{T} - U(T)\ehat{20} (\theta_{0} )\theta_1 =&- \pdiff{\x_0}{t} + \flowVel_c \\
    &+ [P\mat{I}+\eta_3 (T)\vec{B}_0(T,\sin2\theta_0,\cos2\theta_0)] \ehat{10}(\theta_0) \\
    &-\eta_4 (T)\hat{E}^*_{22}(T,\sin2\theta_0,\cos2\theta_0)\hat{\e}_{20}(\theta_0) \\
    &-\eta_2 (T)\hat{E}^*_{12}(T,\sin2\theta_0,\cos2\theta_0)\hat{\e}_3,
\end{split}
\end{equation}
which, after using \cref{xxavgfast}, becomes
\begin{equation}\label{nextord}
\begin{split}
\pdiff{\x_1}{T} - U(T)\ehat{20} (\theta_{0} )\theta_1
 =& - \pdiff{\avg{\x}_0}{t}(t) + \flowVel_c -  \avg{\theta}_{0t} [\Phi_c(T) \ehat{20}(\avg{\theta}_0)-\Phi_s(T) \ehat{10}(\avg{\theta}_0) ] \\
 &+ [P\mat{I}+\eta_3(T)\vec{B}_0(T,\sin2\theta_0,\cos2\theta_0)] \ehat{10}(\theta_0) \\ 
 &-\eta_4 (T)\hat{E}^*_{22}(T,\sin2\theta_0,\cos2\theta_0)\hat{\e}_{20}(\theta_0) \\
 &-\eta_2 (T)\hat{E}^*_{12}(T,\sin2\theta_0,\cos2\theta_0)\hat{\e}_3.
\end{split}
\end{equation}
Here, the $\theta_1$ term on the left-hand side is generated by the next to leading order term in the expansion of $\omega U(T) \ehat{10}(\theta_0)$, and all $\theta$-dependence on the right-hand side reduces to a dependence on $\theta_0=\avg{\theta}_0+\Psi$, as all these contributions are restricted to leading order.

The next step is to apply the Fredholm alternative, though the coupling of the translational dynamics to the angular dynamics entails this is more complicated. Working in the laboratory basis for the spatial components, we write the equations in the matrix form 
\begin{equation}\label{fat}
    \mathcal{L} \begin{bmatrix}
        \x_1\\
        \theta_1
    \end{bmatrix} = \mathcal{G},
\end{equation}
where 
\begin{equation}
    \mathcal{L} = \left[\begin{array}{c|c|c|c}
 \hspace*{-0.2mm} \pdiff{}{T} & 0& 0 & U(T) \sin\theta_0 \\ 
 0 & \pdiff{}{T} & 0 & -U(T) \cos\theta_0 \hspace*{-0.2mm} \\ 
 0 & 0& \pdiff{}{T} & 0\\ 
 0 & 0& 0& \pdiff{}{T}
 \end{array} \right],
\end{equation}
\begin{equation}
    \mathcal{G} = \begin{bmatrix}
        - \diff{\avg{\x}_0}{t}(t) +\flowVel_c -  \avg{\theta}_{0t} [\Phi_c(T) \ehat{20}(\avg{\theta}_0)-\Phi_s(T) \ehat{10}(\avg{\theta}_0) ]+ \cdots \\
        -\avg{\theta}_{0t}+ a_*(T) - b_{**}(T)\sin{2\avg{\theta}_0} -c_{**}(T)\cos{2\avg{\theta}_0}
    \end{bmatrix},
\end{equation}
and $\x_1$ and $\theta_1$ are $2\pi$-periodic in $T$ to match the periodicity of the fast dynamics. Note that some of the terms in $\mathcal{G}$ have been omitted for presentational ease, but are immediately inherited from \cref{nextord}. 
Then, by the Fredholm alternative, non-trivial solutions for $\x_1$ and $\theta_1$ require the forcing $\mathcal{G}$ to be perpendicular to the nullspace of the adjoint of $\mathcal{L}$, that is
\begin{equation}
\mathcal{L}^* = - \left[
 \begin{array}{c|c|c|c}
 \pdiff{}{T} & 0& 0 &0 \\ 
 0 & \pdiff{}{T} & 0 & 0 \\ 
 0 & 0& \pdiff{}{T} & 0\\ 
 -U(T) \sin\theta_0 & U(T) \cos\theta_0 & 0&\pdiff{}{T}
 \end{array} \right],
\end{equation}
with the restriction to $2\pi$-periodic functions in $T$.
Thus, first we determine the null space by solving 
\begin{equation}
    \mathcal{L^*} \vec{s} = \vec{0}
\end{equation}
with the condition of  $2\pi$-periodicity in $T$. This gives
\begin{equation}
    \vec{s} = [
        J_4, J_3, J_2, s_4
    ]\transpose,
\end{equation}
where $J_4$, $J_3$, and $J_2$ are independent of $T$, and thus $2\pi$-periodic in $T$, while $s_4$ satisfies 
\begin{equation}
\begin{split}
    \pdiff{s_4}{T} &=  J_4[U(T)\cos\Psi (T)\sin\avg{\theta}_0 +U(T)\sin\Psi (T)\cos\avg{\theta}_0 ] \\ 
    & -J_3[U(T)\cos\Psi(T) \cos\avg{\theta}_0-U(T)\sin\Psi (T)\sin\avg{\theta}_0], 
\end{split}
\end{equation}
which integrates to
\begin{equation}
s_4 = J_4[U_{cI}(T) \sin\avg{\theta}_0 + U_{sI}(T) \cos\avg{\theta}_0] -J_3[U_{cI}(T) \cos\avg{\theta}_0-U_{sI}(T) \sin\avg{\theta}_0] +J_1,
\end{equation}
with $J_1$ independent of $T$. Noting that both $U(T)$ and $\Psi(T)$ are $2\pi$-periodic in the fast time variable $T$, and recalling that $ \avg{U}_{cI}$, $\avg{U}_{sI}= \bigO{1/\omega}$ from swimmer inefficiency, we have 
\begin{equation}
\begin{split}
    \abs{s_4(T+2\pi)-s_4(T)} &=  \abs{J_4 \sin\avg{\theta}_0 \int_T^{T+2\pi} U(S)\cos\Psi(S) \intd{S} +\cdots} \\  
    &= 2\pi \abs{J_4 \avg{U}_{cI}  \sin\avg{\theta}_0 + \cdots} \\
    &= \bigO{1/\omega} \ll 1, 
\end{split}
\end{equation}
where the additional terms contributing to $s_4$ behave analogously and contribute only at $\bigO{1/\omega}$. As are working at asymptotic accuracy in  $1/\omega \ll 1$, we have  $s_4(T+2\pi)=s_4(T)$ at the current asymptotic order, giving $2\pi$-periodicity in $T$ at the level of accuracy required. With a minor redefinition of $J_1$, such that it can carry a dependence on the slow-time variable $t$, we have that $s_4$ can conveniently be written in terms of $\Phi_s(T)$ and $\Phi_c(T)$ to give the asymptotically accurate null vector via  
\begin{equation}
    \vec{s} =
    \begin{bmatrix}
    J_4 \\ J_3 \\ J_2 \\ J_4 [\Phi_c(T)\sin\avg{\theta}_0+\Phi_s(T)\cos\avg{\theta}_0]  +J_3[ \Phi_s(T)\sin\avg{\theta}_0-\Phi_c(T)\cos\avg{\theta}_0  ]+J_1
 \end{bmatrix}.
\end{equation}
Hence, at the working level of accuracy, the nullspace of $ \mathcal{L^*}$ is spanned by the multipliers of $J_1$, $J_2$, $J_3$, $J_4$ respectively, that is the four linearly independent solutions 
\begin{gather}
\begin{aligned}\label{nsbasis} 
    \vec{p}_1 &= [0,0,0,1]\transpose, & \vec{p}_3 &= [0,1,0,\Phi_s(T)\sin\avg{\theta}_0-\Phi_c(T)\cos\avg{\theta}_0]\transpose,\\
    \vec{p}_2 &= [0,0,1,0]\transpose,
    &\vec{p}_4 &= [1,0,0, \Phi_c(T)\sin\avg{\theta}_0+\Phi_s(T)\cos\avg{\theta}_0]\transpose.
\end{aligned}
\end{gather}
 
The constraint $\inprod{\vec{p}_1}{\mathcal{G}}=0$ with $\inprod{\cdot}{\cdot}$ denoting the inner product immediately generates \cref{fat0} above and, thus, the leading order angular equation of motion of \cref{teqnf}.
With $\avg{z}_0 \coloneqq \e_3\cdot\avg{\x}_0$, imposing $\inprod{\vec{p}_2}{\mathcal{G}}=0$ reveals
\begin{equation}\label{zeqn}
\begin{split}
\diff{\avg{z}_0}{t}(t) = \e_3\cdot\avg{\flowVel_{tr}} &-\avg{\eta_2 (T)\hat{E}^*_{12}(T,\sin2\Psi,\cos2\Psi)}\cos(2\avg{\theta}_0) \\
&+\avg{\eta_2 (T)\hat{E}^*_{11}(T,\sin2\Psi,\cos2\Psi)}\sin(2\avg{\theta}_0),
\end{split}
\end{equation}
the equation of motion for drift perpendicular to the plane of the flow. As can be confirmed below, the location along the $\e_3$ axis, $\avg{z}_0$, does not appear in any of the other equations of motion.

We note that, once more, the rate of strain tensor contributions are taken only after rotating the basis by an angle $\Psi(T)$, induced by the fast rotational dynamics of the swimmer. Additionally, without loss of generality, one may set $\e_3\cdot\avg{ \flowVel_{tr}}$ to zero as, after averaging, it corresponds to a constant velocity since the background flow carries no slow-time dependence. In turn, this constant velocity may be set to zero by the choice of inertial reference frame given the Galilean invariance of the Newtonian mechanics underlying the equations of motion.

Given the above decoupled equation for $\avg{z}_0$ we may, without loss of generality,  overload the symbols $\avg{\x}_0(t)$, $\flowVel_{tr}(T)$, $\mat{\Lambda}(T)$, and $\vec{B}_0$ by considering only their flow-plane components, that is in the $\e_1\e_2$-plane, for the remainder of the analysis. Imposing the constraints $\left\langle\vec{p}_3,\mathcal{G}\right\rangle=\left\langle\vec{p}_4,\mathcal{G}\right\rangle=0$, noting $\avg{\Phi}_c=\avg{\Phi}_s=0$ and with the definition 
\begin{equation}\label{chi} 
\chi =a_*(T)-b_{**}(T)\sin(2\avg{\theta}_0)-c_{**}\cos(2\avg{\theta}_0),
\end{equation} 
so that $\avg{\theta}_{0t}=\avg{\chi}$, 
we arrive at the somewhat lengthy but explicit relation 
\begin{equation} \label{eqnxc000} 
\begin{split}
\diff{\avg{\x}_0}{t}
= & \avg{\flowVel_{tr}}+ \avg{\mat{\Lambda}}\avg{\x}_{0} 
+\avg{\Phi_c\mat{\Lambda}} \ehat{10}(\avg{\theta}_0)
+\avg{\Phi_s\mat{\Lambda}} \ehat{20}(\avg{\theta}_0)
+P( \avg{\cos\Psi} \ehat{10}(\theta_0) +\avg{\sin\Psi} \ehat{20}(\theta_0) )\\  &+\avg{\eta_3(T)\vec{B}_0(T,\sin2\theta_{0},\cos2\theta_{0})\ehat{10}(\theta_{0})} 
-\avg{\eta_4(T)\hat{E}^*_{22} (T,\sin2\theta_{0},\cos2\theta_{0})\ehat{20} (\theta_{0})} \\
& + \avg{\chi \Phi_s }\ehat{10}(\avg{\theta}_0) - \avg{\chi \Phi_c }\ehat{20}(\avg{\theta}_0).
\end{split}
\end{equation}
From this and expanding using the results above, we define the constants $M_1$ $N_1$, $M_2$, $N_2$, which simplify extensively with $\Omega_f(T)=0$ and increasing swimmer symmetry (as explicitly considered in \cref{SymmSimp}) via
\begin{equation}\label{eq: translation before simple notation}
\begin{split}
\diff{\avg{\x}_0}{t}=& \avg{\flowVel_{tr}} + \begin{bmatrix}
    \avg{E^{11}} & \avg{E^{12} -\Omega^*}\\
    \avg{E^{12}+\Omega^*} & -\avg{E^{11}}
\end{bmatrix}\avg{\x}_{0} \\ 
& + \left[P \avg{\cos\Psi} +\avg{\Omega  \Phi_s }\right]\ehat{10}(\avg{\theta}_0)  +\left[P\avg{\sin\Psi} -\avg{\Omega  \Phi_c }\right]\ehat{20}(\avg{\theta}_0) \\  
& + \left[M_1 \sin2 \avg{\theta}_{0} + M_2\cos2 \avg{\theta}_{0}\right] \ehat{10}(\avg{\theta}_0) + \left[N_1 \sin2 \avg{\theta}_{0} + N_2\cos2 \avg{\theta}_{0}\right]\ehat{20}(\avg{\theta}_0).
\end{split}
\end{equation}
Further, we define the linear functions $M$ and $N$ and the constant matrix $\mat{A}$ via
\begin{equation}\label{eqnxc00}
\begin{split}
\diff{\avg{\x}_0}{t}=& \avg{\flowVel_{tr}} + \mat{A} \avg{\x}_0 +\left[P \avg{\cos\Psi} +\avg{\Omega  \Phi_s } +M( \sin2 \avg{\theta}_{0} ,\cos2 \avg{\theta}_{0})\right]\ehat{10}(\avg{\theta}_0) \\
& + \left[ P\avg{\sin\Psi} -\avg{\Omega  \Phi_c } +N( \sin2 \avg{\theta}_{0} ,\cos2 \avg{\theta}_{0})   \right]\ehat{20}(\avg{\theta}_0).
\end{split}
\end{equation}
noting that rates of rotation and shear have no spatial variation for a linear background flow. Deducing the form of each of these expressions requires cumbersome calculation, with significant cancellation needed to arrive at the coefficients $\avg{\Omega  \Phi_s }$ and  $-\avg{\Omega  \Phi_c }$ of  $\ehat{10}(\avg{\theta}_0)$ and $\ehat{20}(\avg{\theta}_0)$, respectively. As with the $\e_3$ direction, we immediately set $\avg{\flowVel_{tr}}=\vec{0}$ without loss of generality by the freedom in the choice of inertial reference frame. 
 
Remarkably, further progress can be made now that the evolution of $\avg{\x}_0$ is written in this form. To proceed, we note that the eigenvalues of $\mat{A}$ are given by 
\begin{equation}
    \pm (\avg{E^{11}}^2+\avg{E^{12}}^2-\avg{\Omega^*}^2)^{1/2}
\end{equation}
and that
\begin{equation}\label{eq: def of Asquared}
    \mat{A}^2 = (\avg{E^{11}}^2+\avg{E^{12}}^2-\avg{\Omega^*}^2) \vec{I},
\end{equation}
as is most readily deduced from the Cayley-Hamilton theorem. Hence, with the definitions 
\begin{equation}
  \nu \coloneqq (\avg{E^{11}}^2+\avg{E^{12}}^2-\avg{\Omega^*}^2)^{1/2}, \quad \mu \coloneqq (\avg{\Omega^*}^2-\avg{E^{11}}^2-\avg{E^{12}}^2)^{1/2},
\end{equation}
we can compute $\exp[\mat{A}t]$ to give
\begin{equation} \label{matexp}
\mat{K}(t)\coloneqq\exp[\mat{A} t]=
\begin{cases}
\cosh(\nu t) \vec{I} + \frac{1}{\nu}\sinh(\nu t) \mat{A},  & \avg{E^{11}}^2+\avg{E^{12}}^2>\avg{\Omega^*}^2, \\
\mat{I} + \mat{A} t, & \avg{E^{11}}^2+\avg{E^{12}}^2=\avg{\Omega^*}^2, \\
\cos(\mu t) \mat{I} + \frac{1}{\mu} \sin(\mu t) \mat{A}, &  \avg{E^{11}}^2+\avg{E^{12}}^2<\avg{\Omega^*}^2.
\end{cases}
\end{equation} 
Thus, solving \cref{eqnxc00} in terms of $\mat{K}$ and its convolution reveals 
\begin{multline}\label{solxc00}
    \avg{\x}_{0} (t) = \mat{K}(t) \avg{\x}_{0}(t=0) \\ 
    +\int_0^t \mat{K} (t-s) \vec{C}( \sin2 \avg{\theta}_{0} (s),\cos2 \avg{\theta}_{0}(s), \sin \avg{\theta}_{0} (s),\cos \avg{\theta}_{0}(s))\intd{s},
\end{multline}
where we define
\begin{equation}\label{eq: C def}
\begin{split}
\vec{C} &\coloneqq \left[P \avg{\cos\Psi} +\avg{\Omega  \Phi_s } +M( \sin2 \avg{\theta}_{0} ,\cos2 \avg{\theta}_{0})\right]\ehat{10}(\avg{\theta}_0) \\ 
& + \left[ P\avg{\sin\Psi} -\avg{\Omega \Phi_c } +N( \sin2 \avg{\theta}_{0} ,\cos2 \avg{\theta}_{0})\right]\ehat{20}(\avg{\theta}_0),
\end{split}
\end{equation} 
with $M$ and $N$ linear in their arguments. 

\section{Classification of rotational dynamics}\label{sec: features: rotation}
Numerous deductions can be made from both the leading order multiple scales solution for $\theta$ in \cref{tsol} and for $\x_{0}(t,T)$ and $\avg{\x}_{0} (t)$ in \cref{xc00,zeqn,matexp,solxc00}. Such conclusions concern whether the swimmer rotational dynamics asymptotes to rocking, tumbling or a steady angle, and whether there is inexorable drift or oscillation in the translational dynamics.

The rotational dynamics for a rapidly deforming planar swimmer within a planar linear flow, with a possible fast oscillation, is given by \cref{tsol}. If we assume that $a^2>b^2+c^2$, we have 
\begin{equation} \label{teqn000}
\avg{\theta}_0 = \arctan\left(\frac{q\tan\theta_{00} + [a-c-b\tan\theta_{00}]\tan(qt)}{q+[b-(a+c)\tan\theta_{00}]\tan(qt)}\right),
\end{equation}
where $q=(a^2-b^2-c^2)^{1/2}$. Note that there is a potential degenerate edge case with 
\begin{equation}
    a-c-b\tan\theta_{00}= 0 = b-(a+c) \tan\theta_{00}, 
\end{equation}
which would give $\avg{\theta}_0=\theta_{00}$ for all time. However, eliminating $\tan\theta_{00}$ in favour of $a,b,c$ immediately yields $a^2=b^2+c^2$, violating our assumption. Thus, this degeneracy cannot be realised. Similarly, other degenerate cases lie out of reach, such as setting the numerator equal to zero in \cref{teqn000}, which requires $\tan\theta_{00}=0$ and $a=c$, once more violating $a^2>b^2+c^2$.

This rotational dynamics corresponds to a continuous tumble (rather than rocking back and forth). To see this, note that within the expression for $\avg{\theta}_0$ given by \cref{teqn000}, we have $\tan(qt)$ increasing in time monotonically, and the right-hand side is monotonic in $\tan(qt)$ (noting that we have excluded degenerate cases where the expression is constant). Thus, the jump in $\arctan$ to maintain continuity as $qt$ passes though $\pi/2+n\pi$ for some integer $n$ is always in the same direction, so that $\avg{\theta}_0$ changes monotonically with increasing time. The non-dimensional period of the tumbling is given by 
\begin{equation} \label{period}
    \frac {2\pi} q, 
\end{equation}
with $q=(a^2-b^2-c^2)^{1/2}$ once more. The factor of two in the numerator arises because each jump to a new branch of $\arctan$ results in an increase in $\avg{\theta}_0$ by $\pi$ once $qt$ propagates across the new branch, so that propagation across two branches is required to increase $\avg{\theta}_0$ by a full rotation of $2\pi$.

With this, and noting the long-time limits of the other cases of \cref{tsol}, we thus have a necessary and sufficient condition for the swimmers of \cref{Ges} (with the symmetries of \cref{geshs}) to tumble. In particular, endless tumbling is guaranteed precisely when $a^2>b^2+c^2$, that is 
\begin{align}\label{ac1}
    \left[\avg{\flowAngSpeed(T)+\Omega(T)}\right]^2 \nonumber\\ 
    > &\left[  \avg{B(T)\hat{E}^*_{11}(T,\sin2\Psi,\cos2\Psi)-\lambda_5(T)\hat{E}^*_{12}(T,\sin2\Psi,\cos2\Psi)}\right]^2 \nonumber\\ 
    + &\left[\avg{B(T)\hat{E}^*_{12}(T,\sin2\Psi,\cos2\Psi)+\lambda_5(T)\hat{E}^*_{11}(T,\sin2\Psi,\cos2\Psi)}\right]^2.
\end{align} 
For all bodies of \cref{geshs}, except those that possess only the $C_3$ symmetry within the gait cycle, this reduces to 
\begin{multline}\label{ac2}
    \left[\avg{\flowAngSpeed(T)+\Omega(T)}\right]^2 \\ 
    >\left[\avg{B(T)\hat{E}^*_{11}(T,\sin2\Psi,\cos2\Psi) }\right]^2+\left[\avg{B(T)\hat{E}^*_{12}(T,\sin2\Psi,\cos2\Psi)}\right]^2.
\end{multline}
For any body of \cref{geshs} that also possesses fore-aft symmetry throughout the gait cycle, the condition further reduces to
\begin{equation}\label{ac3} 
    \left[\avg{\flowAngSpeed(T) }\right]^2 > \left[\avg{B(T)E_{11}(T)}\right]^2 + \left[\avg{B(T)E_{12}(T)}\right]^2.
\end{equation}
This latter relation readily collapses onto $B^2<1$ if the particle is rigid and the flow is taken to be a simple, time-independent shear, in line with the classical results of \citet{Jeffery1922} and \citet{Bretherton1962}. Notably, none of these criteria depend in any way on the translational swimming motility, with no dependence on $P$ nor $U_I$.

Each of these conditions signifies that tumbling occurs once the angular forcing  
\begin{equation}
    \left[\avg{\flowAngSpeed(T)+\Omega(T)}\right]^2
\end{equation} is sufficiently high, where the threshold for tumbling depends on the details of the interactions between the deformation of the swimmer and the rate of shear experienced by the swimmer, accounting for any fast-timescale changes in its orientation via the angle $\Psi(T)$. 
Notably, rocking \emph{never} occurs in the swimmer system: in every case of \cref{tsol}, the swimmer angle never oscillates back and forth without whole turns. Instead, we have that the swimmer either tumbles or its orientation asymptotes to a fixed angle. This is in distinct contrast to the behaviour of a simple pendulum, where whole turns are replaced by rocking as the forcing is reduced.

Furthermore, if the swimmer is such that the tumbling condition is given by \cref{ac3}, we see that increasingly elongated swimmers (i.e. those with larger $B(T)$) are always less prone to tumbling than less-elongated swimmers, so long as the flows are such that $E_1(T)$ and $E_2(T)$ do not change sign. In other words, the more elongated the swimmer, the more that tumbling is suppressed in this setting. Further, the period of tumbling $2\pi/q$, increases with tumbling, approaching infinity as we leave this dynamical regime and $q$ approaches zero. 

However, this simple conclusion need not hold in more generality. For instance, should $E_1(T)$ or $E_2(T)$ change sign over a period, then there is no such guarantee on the scaling (or indeed whether there is an increase or decrease) of $b^2+c^2$ in \cref{ac3} as $B(T)$ increases in magnitude. In particular, such time-dependent details can drive the system into a regime where tumbling is suppressed. This, along with other similarly complicating factors like fast swimmer oscillations, exemplifies and emphasises the more general observation, as also noted in previous studies \citep{Walker2022,Walker2023}, that simply using averaged parameters for flow and swimmer properties can generate fundamentally different and incorrect predictions. Ultimately, this is simply because the operations of averaging and multiplication do not commute.

\section{Classification of translational dynamics}\label{sec: features: translation}
From \cref{xxavgfast}, we have that the trajectory averaged over the fast oscillations, $\bar{\x}_0(t)$, is perturbed by a fast oscillation of zero mean at the leading order of the multiple scales approximation. Furthermore, by inspection of \cref{zeqn}, any drift in the $\e_3$ direction perpendicular to the plane of the flow can be treated independently once $\avg{\theta}_0$ is known. In addition, this drift is completely decoupled and only driven by the background flow, unless the swimmer only possesses a $D_n, n\geq 4$ symmetry for part of its gait cycle. In the latter case, shape changes in the body, encapsulated by $\eta_2(T)$, can interact with the strain rate of the flow to generate a non-trivial drift perpendicular to the flow, even in the absence of a background flow component in this direction. In contrast, the dynamics for $\avg{\x}_0(t)$ in the plane of flow is much more complex, even for swimmers with high symmetry, which we consider below. 

\subsection{Exponential temporal dynamics in the plane}
If the average strain rate of the flow  dominates the average rotation rate of the flow, such that
\begin{equation}
    \avg{E^{11}}^2+\avg{E^{12}}^2 > \avg{\Omega^*}^2,
\end{equation}
then the matrix exponential of \cref{matexp} entails that the swimmer will drift away from its starting point at an exponential rate, irrespective of its angular dynamics and with the possible exception of edge cases. Such edge cases can occur when $\mat{K}(t) \avg{\x}_{0}(t=0)$ is precisely balanced by 
\begin{equation}\label{treq2} 
    \int_0^t \mat{K} (t-s) \vec{C}( \sin2 \avg{\theta}_{0} (s),\cos2 \avg{\theta}_{0}(s), \sin \avg{\theta}_{0} (s),\cos \avg{\theta}_{0}(s))\intd{s}.
\end{equation}
However, \cref{treq2} is independent of the initial location of the swimmer. Thus, mathematical precision is required in the initial conditions for such an edge case, which would not be realisable in practice.

\subsection{Linear temporal dynamics in the plane}
We proceed to consider the degenerate case 
\begin{equation}
    \avg{E^{11}}^2+\avg{E^{12}}^2 = \avg{\Omega^*}^2,
\end{equation}
where $\mat{A}^2=0$ from \cref{eq: def of Asquared}. This splits into two further subcases: $\mat{A}=\vec{0}$ or $\mat{A}\neq\vec{0}$. 

If $\mat{A}=\vec{0}$, the flow is either trivial or it is oscillatory with zero mean, so that $\avg{E^{11}}=\avg{E^{12}}= \avg{\Omega^*}=0$. If the flow is trivial, the equations for translation collapse to 
\begin{equation}
    \frac{\mathrm{d} \avg{\x}_{0}} {\mathrm{d}t} = P\avg{\cos\Psi}\ehat{10}(\avg{\theta}_0)+ P\avg{\sin\Psi}\ehat{20}(\avg{\theta}_0) + \avg{\Omega\Phi_s}\ehat{10}(\avg{\theta}_0) - \avg{\Omega\Phi_c}\ehat{20}(\avg{\theta}_0),
\end{equation} 
and  the swimmer  progresses, in general, on a curved trajectory. However, its dynamics is modulated by the fast rotation of the swimmer, $\Omega_{f}(T)$ through $\Psi,\Phi_s,\Phi_c$, albeit in a more complex manner than suggested by naive averaging, with $\avg{\Omega}_f=0$ insufficient to simplify further. It is useful to note that $P=0$ is insufficient to guarantee the absence of progressive swimming in a quiescent fluid with non-progressive swimming requiring the additional constraints $\avg{\Omega\Phi_s}=\avg{\Omega\Phi_c}=0$. Further, if the swimmer has no fast rotation at all, so that $\Omega_f=0$, then $\Psi=\Phi_s=0$  and we have 
\begin{equation}
    \diff{\avg{\x}_0}{t} = P \ehat{10}(\avg{\theta}_0)
     - \avg{\Omega\Phi_c}\ehat{20}(\avg{\theta}_0)
\end{equation}
and, hence, the swimming corresponds to a curved trajectory with radial velocity $P$ and a velocity in the $\avg{\theta}_0$ direction of $-\avg{\Omega\Phi_c}$.

If instead we have $\mat{A} = \vec{0}$ via a non-trivial mean-zero oscillatory flow, we have $\mat{K} = \mat{I}$ and 
\begin{equation}\label{eq: translation: nontrivial oscillatory}
    \avg{\x}_0(t)= \avg{\x}_0(t=0)+\int_0^t \vec{C}(\sin2\avg{\theta}_{0}(s),\cos2\avg{\theta}_{0}(s), \sin\avg{\theta}_{0} (s),\cos\avg{\theta}_{0}(s))\intd{s},
\end{equation}
with $\vec{C}$ as in \cref{eq: C def}. Suppose that we are in one of the regimes explored in \cref{sec: features: rotation} in which the orientation of the swimmer asymptotes to a fixed value. Then, the integrand of \cref{eq: translation: nontrivial oscillatory} becomes constant for large values of the integration variable and, thus, the swimmer will drift to infinity linearly in time (neglecting possible edge cases of perfect cancellation between terms). Notably, this linear drift can happen even if there is only reciprocal swimming, i.e. $P=\avg{\Omega\Phi_s}=\avg{\Omega\Phi_c}=0$. For instance, in the simple setting of fore-aft symmetry and a body of revolution, so that the equations reduce to those of \cref{eqnxc000app1} in \cref{SymmSimp}, there is a contribution  
\begin{equation}\label{UE12} 
  \avg{U_IE^{12}} \left[\sin2\avg{\theta}_0\ehat{10}(\avg{\theta}_0) + \cos2\avg{\theta}_0\ehat{20}(\avg{\theta}_0)\right]
\end{equation}
in the plane of the flow, which ultimately arises from the simplification of the $\avg{\Phi_c\mat{\Lambda}}\ehat{10}   (\avg{\theta}_{0})$ term. This contribution need not be zero even if
\begin{equation}
    P=\avg{\Omega\Phi_s}=\avg{\Omega\Phi_c}=\avg{U_I}=\avg{E^{11}}=\avg{E^{12}}=\avg{\Omega^*}=0.
\end{equation}
Thus, seemingly non-progressive swimming can generate a drift to spatial infinity through interactions with purely oscillatory, mean-zero flows. In other words, this provides an explicit mechanism by which a swimmer might circumvent Purcell's scallop theorem.

Suppose instead that we are in a regime in which the swimmer tumbles endlessly in the zero-mean oscillatory flow. Note that these flows necessarily have $\avg{\Omega^*}=0$, so that swimmer tumbling requires the absence of fore-aft symmetry in order to be admissible (consider \cref{ac3} with $\avg{\Omega^*}=0$). Even without this symmetry, we require $\avg{\Omega}^2>0$ in order for \cref{ac1} and \cref{ac2} to admit any solutions. In this case, we can evaluate the contribution of the term associated with $P\avg{\cos\Psi}$ in the $\e_1$ direction in \cref{eq: translation: nontrivial oscillatory} over a single period of rotation. For one such tumble, which we recall has period $2\pi/q$, starting from some $t=t_s$, this contribution is $P\avg{\cos\Psi}$ multiplied by 
\begin{equation}\label{eq59}
\begin{split}
    \int_{t_s}^{t_s+2\pi/q} \cos\avg{\theta}_0 \mathrm{d}t &= \int_{\avg{\theta}_0(t_s)}^{\avg{\theta}_0(t_s)+2\pi} \frac{\cos\avg{\theta}_0}{a-b\sin2\avg{\theta}_0-c\cos2\avg{\theta}_0} \intd{\avg{\theta}_0} \\ 
    &=\frac{1}{a}\int_{-\pi}^{\pi} \frac{\cos\avg{\theta}_0}{1-R\cos(2\avg{\theta}_0-2\nu)} \intd{\avg{\theta}_0}, \text{ with } R = \frac{(b^2+c^2)^{1/2}}a < 1
\end{split}
\end{equation} 
and $\nu$ a phase shift. Here, we have recalled the governing equation of \cref{teqnf} to change variables to $\avg{\theta}_0$ in the integrals. Using the periodicity of the cosines, this can be written as a linear combination of the following integrals: 
\begin{equation}\label{eq510}
    \frac{1}{a} \int_{-\pi}^{ \pi} \frac{\cos\avg{\theta}_0}{1 -R\cos(2\avg{\theta}_0)} \intd{\avg{\theta}_0}=\frac{2}{a} \int_{0}^{\pi} \frac{\cos\avg{\theta}_0}{1-R\cos(2\avg{\theta}_0)} \intd{\avg{\theta}_0}, \quad \frac{1}{a}\int_{-\pi}^{ \pi} \frac{\sin\avg{\theta}_0}{1-R\cos(2\avg{\theta}_0)} \intd{\avg{\theta}_0}.
\end{equation}
Both of these integrals are zero, by the odd parity in reflection about $\theta_0=\pi/2$ for the first integral and about $\theta_0=0$ for the second. Similarly, all other terms contributing to the translational motility in \cref{eq: translation: nontrivial oscillatory} can be written as a linear combination of integrals over a temporal period, with integrands 
\begin{equation}
    \sin\theta_0, \quad \sin2\theta_0\cos\theta_0, \quad\cos2\theta_0\cos\theta_0, \quad\sin2\theta_0\sin\theta_0,\quad\sin2\theta_0\sin\theta_0 ,
\end{equation}
which all integrate to zero using the same arguments as above. Hence, we may conclude that \emph{a tumbling swimmer in a purely oscillatory linear planar flow does not drift.} 

Now we consider the final subcase of linear temporal dynamics in the plane, assuming that $\mat{A}\neq\vec{0}$ and $\mat{A}^2 = \vec{0}$. We consider the Jordan normal form for $\mat{A}$ to within an overall scaling, though it is also useful to explicitly demonstrate that the transformation required in this particular case is a rotation.  First, let $\vec{e}_A$ denote the zero-eigenvalue unit eigenvector of $\mat{A}$ (unique up to sign) and, thus, $\mat{A}\vec{e}_A= \vec{0}$. Additionally, let $\vec{e}_A^\perp$ denote the unit vector perpendicular to $\vec{e}_A$ so that $\{\vec{e}_A,\vec{e}_A^\perp\}$ is a right-handed orthonormal basis. We have 
$\mat{A}\vec{e}_A^\perp\neq \vec{0} $, otherwise $\mat{A}=\vec{0}$. Then, 
with $\alpha_1$ and  $\alpha_2$ defined by 
\begin{equation}
    \mat{A}\vec{e}_A^\perp = \alpha_1\vec{e}_A+\alpha_2 \vec{e}_A^\perp
\end{equation}
we have $\vec{0} = \mat{A}^2\vec{e}_A^\perp = \alpha_2  \mat{A}\vec{e}_A^\perp$
and, hence, $\alpha_2=0$ and $\alpha_1\neq 0$. Thus, 
\begin{equation}
    \mat{A} \left[\left. \vec{e}_A \right| \vec{e}_A^\perp \right] =  \left[\left. \vec{0}\right| \alpha_1 \vec{e}_A  \right] = \left[\left. \vec{e}_A \right| \vec{e}_A^\perp \right]
    \begin{bmatrix} 
    0 & \alpha_1 \\
    0 & 0 
    \end{bmatrix}.
\end{equation}
Noting that $ \left[\left. \vec{e}_A \right| \vec{e}_A^\perp \right]$ is an orthogonal matrix, using its transpose to left multiply both sides 
shows that a  rotation of the axes can be found to transform $\mat{A}$ to a matrix that is zero except for the upper right off-diagonal entry (this is the Jordan normal form of $\mat{A}$ to within scaling). Hence, the flows we are considering here  are, on averaging, those of pure shear with $\avg{\flowVel} =2\avg{E^{12}} y\vec{e}_1$ for a suitable choice of orthonormal basis, and we have $\avg{E^{12}}\neq 0$ as $\mat{A}\neq\vec{0}$.

For swimmers that asymptote to a fixed angle, the presence of pure shear results in infinite drift in the $\e_1$ direction that increases quadratically in time in general. This arises from the fact that the dominant term in the $\e_1$ direction for $t\gg t_*\gg 1$ (sufficiently large) is approximately 
\begin{multline} 
    \left[ \int_{t_*}^{t} \mat{I} +\mat{A}(t-s) \intd{s}\right]\left.\vec{C}( \sin2 \avg{\theta}_{0} ,\cos2 \avg{\theta}_{0}, \sin \avg{\theta}_{0} ,\cos \avg{\theta}_{0})\right\rvert_{\avg{\theta}_{0}=\avg{\theta}_0(\infty)} \\ 
    = \left[ \int_{t_*}^{t} 
    \begin{pmatrix} 
        1 & 2E^{12} (t-s) \\ 
        0 & 1 
    \end{pmatrix}\intd{s}\right] \left.\vec{C}( \sin2 \avg{\theta}_{0} ,\cos2 \avg{\theta}_{0}, \sin \avg{\theta}_{0} ,\cos \avg{\theta}_{0})\right|_{\avg{\theta}_{0}=\avg{\theta}_0(\infty)}  \\
    \sim \begin{pmatrix}
            \ord{t^2} \\ 
            \ord{t} 
        \end{pmatrix}.
\end{multline} 
More generally, movement in the $\e_2$ direction is simply inherited from the case $\mat{A}=\vec{0}$, as 
$ \mat{K} = \mat{I}+t\mat{A}$ and $\mat{A}$ will not generate contributions along the $\e_2$ direction (its second row is zero). Thus, the drift in this direction is linear in time. Notably, this drift to infinity in the $\e_2$ direction (and thus across pathlines), can still occur even if there is no net swimming in the absence of a background flow, i.e. whenever
\begin{equation}
    P=\avg{\Omega\Phi_s}=\avg{\Omega\Phi_c}=0.
\end{equation} In particular, in the simple case of fore-aft symmetry and a body of revolution (summarised by \cref{eqnxc000app1}) there is a contribution in the $\e_2$ direction of the form $\avg{U_I\Omega}\cos\avg{\theta}_0$ that will induce this drift. This, in turn, again demonstrates that swimmer interactions with flow can break Purcell's scallop theorem. Further, in this example the background flow need not be oscillatory, as the drift is driven by interaction between the swimmer's translation and oscillation, with $\avg{U_I\Omega}\neq 0$ a possibility even if $\avg{U_I}=\avg{\Omega}=0$. 

Finally, we note that a tumbling swimmer's trajectory will not drift in the $\e_2$ direction. In particular, given the swimmer is tumbling and noting $\e_2 \transpose \vec{A}=\vec{0}$, we have that its location 
in the $\e_2$ direction is given by 
\begin{subequations}
\begin{align} \e_2\transpose  \avg{\x}_0(t) &=   \e_2\transpose \avg{\x}_0(t=0)\nonumber \\ &\quad +  \e_2\transpose \int_0^t(\mat{I} +\mat{A}(t-s)   )\vec{C}(\sin2\avg{\theta}_{0}(s),\cos2\avg{\theta}_{0}(s), \sin\avg{\theta}_{0} (s),\cos\avg{\theta}_{0}(s))\intd{s}, \\ &=    \e_2\transpose \avg{\x}_0(t=0)+\int_0^t  \e_2\transpose\vec{C}(\sin2\avg{\theta}_{0}(s),\cos2\avg{\theta}_{0}(s), \sin\avg{\theta}_{0} (s),\cos\avg{\theta}_{0}(s))\intd{s}.
\end{align}
\end{subequations}
The integral above is the second component of those occurring in \cref{eq: translation: nontrivial oscillatory} for a tumbling swimmer, as considered in \cref{eq59,eq510}, and, thus, by an inheritance of this analysis we can deduce that there is no drift in the $\e_2$ direction for a tumbling swimmer in a flow that is pure shear and non-trivial after time averaging.

\subsection{Oscillatory temporal dynamics in the plane}
The final class of translational dynamics to consider occurs if the average strain rate is dominated by the rotation rate, such that 
\begin{equation}
    \avg{E^{11}}^2+\avg{E^{12}}^2 <\avg{\Omega^*}^2.
\end{equation}
Neglecting edge cases, we first consider swimming in which $a^2 < b^2+c^2$, so that there is no tumbling and hence, the swimmer tends to a fixed angle for large time (see \cref{sec: features: rotation}). Noting from \cref{matexp} that $\mat{K}(t)$ is periodic on the slow timescale with period $2\pi/\mu$, where
\begin{equation}
    \mu = (\avg{\Omega^*}^2-\avg{E^{11}}^2-\avg{E^{12}}^2)^{1/2},
\end{equation}
we have for $t_*$ sufficiently large
\begin{multline} 
 \avg{\x}_0(t_*+2\pi/\mu)- \avg{\x}_0(t_*) \\
 \approx \left[ \int_{t_*}^{t_*+2\pi/\mu}\mat{K}(t_*-s)\intd{s}\right] \left.\vec{C}( \sin2 \avg{\theta}_{0} ,\cos2 \avg{\theta}_{0}, \sin \avg{\theta}_{0} ,\cos \avg{\theta}_{0})\right|_{\avg{\theta}_0=\avg{\theta}_0(\infty)}=0.
\end{multline}
The error in approximating $\theta_0$ by its asymptote is exponentially small for large time (see \cref{tsol}) and, hence, these errors do not accumulate. Thus, the trajectory is asymptotically periodic and bounded for large time.
Furthermore, even with net motion of the swimmer, so that at least one of $P$, $\avg{\Omega\Phi_c}$, or  $\avg{\Omega\Phi_c}$ is not zero, the contribution of the progressive swimming to the motion for $t>t_*$ (with $t_*$ sufficiently large) is zero. For example, when $P>0$, the contribution to the motion that scales with $P\avg{\cos\Psi}$ is (to within exponentially small errors) given by 
\begin{multline}
    P \avg{\cos\Psi} \int_{t_*}^t \mat{K} (t_*-s)  \ehat{10}(\infty)\intd{s} \\ 
    = P\avg{\cos\Psi}\left( \frac {\sin(\mu t)-\sin(\mu t_*)}\mu \ehat{10}(\infty) + \frac {\cos(\mu t_*)-\cos(\mu t)}{\mu^2}\mat{A} \ehat{10}(\infty) \right)
\end{multline}
with $\mat{K}$ as in \cref{matexp}. In particular, this contribution is oscillatory. Directly analogous and oscillatory results also hold for all terms involving $P$, $\avg{\Omega\Phi_c}$ and  $\avg{\Omega\Phi_s}$. In turn, this explicitly demonstrates that the progressive movement of the swimmer has been converted to an oscillatory movement by the background flow.

In contrast, if the tumbling condition holds, that is $a^2 > b^2 + c^2$,  the dynamics will involve the convolution of oscillations at the tumbling frequency and at the frequency associated with the flow 
\begin{equation}
 \mu = (\avg{\Omega^*}^2-\avg{E^{11}}^2-\avg{E^{12}}^2)^{1/2}.
\end{equation}
Whether such dynamics induces oscillations or unbounded dynamics in the trajectory at long time is contingent on whether or not there is resonance between the different oscillatory contributions, though oscillations may be typically expected as resonance requires parameter fine-tuning. Though general results in this case are less forthcoming, we can examine special cases of \cref{tsol} and \cref{solxc00} to yield additional insights, as we pursue below. 

\section{Special cases of fore-aft symmetric swimmers}
\label{sec: special cases}
While the analytical cases above considered relatively general bodies, emphasising the ubiquity of the observations among different body shapes, the special cases below are restricted to bodies of revolution that possess fore-aft symmetry, unless explicitly stated otherwise. Hence, only $-\lambda_2(T)\equiv B(T)$, the Bretherton parameter, is non-zero for the interactions between the rate of strain and the swimmer dynamics in \cref{decomp}. In addition, fore-aft symmetry implies that $\Omega_f=\Omega=0$ and, hence,
\begin{equation}
    U_{sI}=0,  \quad \Phi_s(T)=0, \quad U_I(T)=U_{cI}(T), \quad \avg{U}_{cI}=0,\quad \Phi_c(T) = U_I(T).
\end{equation}
With these restrictions, the angular dynamics are given by \cref{tsol} with the additional reductions 
$ a_*=\Omega^*$, $b_*=BE^{11}$, and $c_*=-BE^{12}$, which arise from \cref{abc,cstarstar,teqnf,bcsimp}.
For clarity and the interested reader, the equations of motion are explicitly simplified in \cref{SymmSimp}, where we have from \cref{eqnxc000app1} that motion in the $\e_3$ direction is trivial. Thus, the translational equation of motion in the flow plane reduces to
\begin{multline}\label{eqnxc00s} 
\diff{\avg{\x}_0}{t} = \underbrace{\begin{pmatrix}
\avg{E^{11}} & \avg{E^{12} -\Omega^*} \\
\avg{E^{12}+\Omega^*} & -\avg{E^{11} }
\end{pmatrix}}_{\mat{A}} \avg{\x}_{0} + 
\begin{pmatrix}
P+\avg{U_I E^{11}} & \avg{U_I ( E^{12} -\Omega^*)} \\ 
\avg{U_I ( E^{12} +\Omega^*)} & P -\avg{U_I E^{11}} 
\end{pmatrix}\ehat{10}(\avg{\theta}_0 )\\ + 
 \left(\avg{U_I BE^{11}}\sin 2\avg{\theta}_0-\avg{U_I BE^{12}}\cos 2\avg{\theta}_0 -\avg{U_I\Omega^*}\right)\ehat{20}(\avg{\theta}_0 ),
\end{multline} 
dropping the constant background flow contribution, $\avg{\vec{u}}_{tr}^*$, without loss of generality by the choice of inertial reference frame.

In this reduced yet still complex setting, we will explore swimmer behaviours in various planar linear flows and demonstrate that simple dynamics emerge despite the level of complexity remaining.

\subsection{Rotational flow}
A particularly simple case is that of rotational flow, for which $E^{11}=E^{12}=0$ and $\avg{\Omega^*} \neq 0$. In this regime, the tumbling condition of \cref{sec: features: rotation} holds, with 
$a= \avg{\Omega^*}\neq 0$ and $b=c=0$, so that \cref{tsol} reduces to
\begin{equation}\label{eq: rotational flow: angle}
   \avg{\theta}_0 = \theta_{00}+\avg{\Omega^*}t, 
\end{equation}
where $\theta_{00}=\avg{\theta}_0(t=0)$.
Further, with $\avg{x}_0=\e_1\cdot\avg{\x}_0$, $\avg{y}_0=\e_2\cdot\avg{\x}_0$, and noting that the $\avg{U_I\Omega^*}$ terms precisely cancel in \cref{eqnxc00s}, we have
\begin{equation}
    \diff{\avg{x}_{0}}{t} = - \avg{\Omega^*}\avg{y}_0 +P\cos\avg{\theta}_0, \quad 
    \diff{\avg{y}_{0}}{t} = \avg{\Omega^*}\avg{x}_0 +P\sin\avg{\theta}_0.
\end{equation}
Hence, if there is no intrinsic net swimming ($P=0$), the overall motion is that of simple harmonic oscillations, with 
\begin{equation}
    \diff{^2\avg{x}_0}{t^2} + \avg{\Omega^*}^2 \avg{x}_0 = 0
\end{equation}
and, thus, there is no net motion on the long timescale for non-progressive swimmers in purely rotational flow. 

For $P\neq 0$, we instead have the dynamics of a forced oscillator, with
\begin{equation}\label{eq: rotational flow: forced oscillator}
\diff{^2\avg{x}_0}{t^2} + \avg{\Omega^*}^2 \avg{x}_{0}
= -2Pa \sin\left(\theta_{00}+\avg{\Omega^*}t\right), \quad
 \avg{y}_{0} = \frac{1}{\avg{\Omega^*}}\left[P\cos\avg{\theta}_0 - \diff{\avg{x}_{0}}{t}\right],
\end{equation} 
which generates resonance with no parameter fine-tuning beyond that needed to force a purely rotational flow. Thus, for purely rotational flow and even slightly progressive swimmers, we conclude that swimmer motility generates a resonant oscillation, so that the swimmer distance from the origin scales linearly with time.
Recovering additional generality by reinstating the shape parameters $\lambda_5$, $\eta_2$, $\eta_3$ and $\eta_4$ does not change these observations, as they enter the equations of motion through the rate of strain tensor, which here is zero. 

However, if we relax the constraint of fore-aft symmetry and allow the swimmer to generate a slow-timescale rotation $\Omega\neq0$, then $a = \avg{\Omega} +\avg{\Omega^*} \neq \avg{\Omega^*}$. In turn, \cref{eq: rotational flow: angle} no longer holds and the corresponding forcing in \cref{eq: rotational flow: forced oscillator} is modified to $-2Pa\sin\left(\theta_{00} + at\right)$. Thus, the swimmer oscillates and resonance does not occur, demonstrating that some degree of fine tuning (i.e. fore-aft symmetry) is indeed required for resonance.

\subsection{Irrotational flow}
For a non-trivial irrotational flow, we have $\Omega^*=0$ and 
$\avg{E^{11}}^2 +\avg{E^{12}}^2>0$. Thus, we also have $a=0$ and, by \cref{sec: features: rotation}, $\avg{\theta}_0$ tends to a constant asymptote for large time, and we see exponential dynamics in the plane. Excluding the edge case noted in \cref{sec: features: rotation}, the swimmer drifts off to spatial infinity. This long-time trajectory can be examined in more detail without further loss of generality. As $\Omega^*=0$, the matrix $\mat{A}$ of \cref{eqnxc00s} is symmetric and, hence, diagonalisable. Thus, with a suitable choice of laboratory basis, the equations of motion in the plane of the flow in the long-time limit are given approximately by
\begin{equation}\label{eq: irrotational: long-term trajectory}
\diff{\avg{x}_{0}}{t} = \left(\avg{E^{11}}^2 +\avg{E^{12}}^2\right)^{1/2} ( \avg{x}_{0}+\alpha), \quad
\diff{\avg{y}_{0}}{t}= -\left(\avg{E^{11}}^2 +\avg{E^{12}}^2\right)^{1/2} ( \avg{y}_{0}+\beta),
\end{equation}
where $\alpha$ and $\beta$ are the long-time limits of the other terms in \cref{eqnxc00s}. Noting the opposing signs of the eigenvalues in \cref{eq: irrotational: long-term trajectory}, this explicitly demonstrates an exponential drift to infinity in one direction accompanied by an exponential decay towards a constant in the orthogonal direction. Further, we have
\begin{equation}
    \diff{\avg{y}_{0}}{\avg{x}_{0}} = 
-\frac{ \avg{y}_{0}+\beta }{ \avg{x}_{0}+\alpha} 
\end{equation}
and, hence,
\begin{equation}\label{hyp}
    y_0+\beta =\frac{M}{\avg{x}_{0}+\alpha}, 
\end{equation}
where $M$ is a constant of integration. Thus, independent of any further details, swimmers in such irrotational flows move along hyperbolae in the plane. 

Notably, these broad conclusions apply for both progressive and non-progressive swimmers. They also apply if we relax many of our symmetry constraints (with the exception of fore-aft reflection invariance), with these results holding for non-zero $\eta_2$, $\eta_3$, $\eta_4$, and $\lambda_5$. The breaking of fore-aft symmetry, however, allows for self-induced rotation that has the potential to invalidate these conclusions, as we plausibly obtain $a>0$ and lose the asymptoting behaviour of the swimmer orientation.
 
\subsection{Motility in stationary shear}
The classical example of a stationary shear flow can be recovered by setting $E^{11}=0$ and $\Omega^*=-E^{12}\neq0$, without loss of generality. In this case, \cref{tsol,solxc00} entail that $\ybar$ and $\thetabar$ decouple from $\avg{x}_0$, with motion in the latter direction given by an inexorable drift in all but edge cases. Hence, we focus on the dynamics of $\ybar$ and $\thetabar$, which here are governed by the reduced system
\begin{subequations}
\begin{align}
    \diff{\ybar}{t} &= -E^{12}\avg{\UI\B}\cos{\thetabar}\cos{2\thetabar} +P\sin{\thetabar},\\
 \diff{\thetabar}{t} &= -E^{12}\left(1 - \avg{B}\cos{2\thetabar}\right).\label{eq: stationary shear: angle}
\end{align}
\end{subequations}

We first consider the case with $\abs{\avg{B}}<1$. \Cref{eq: stationary shear: angle} immediately implies that $\thetabar$ is periodic, which can also be seen in the general formalism of \cref{teqnf,tsol} by setting $a=-E^{12}$, $b=0$, $c= -E^{12} \avg {B}$, and $q=(a^2-b^2-c^2)^{1/2}=\abs{E^{12}}(1-\avg {B}^2)^{1/2}>0$. This dynamics corresponds precisely to that of planar Jeffery's orbits \citep{Jeffery1922}, as generalised by Bretherton \citep{Bretherton1962} and identified in planar shape-changing swimmers by \citet{Gaffney2022b}.

We also have that $\ybar$ is periodic, as can be deduced by observing that the system is conservative, with
\begin{equation}
 \diff{H}{t}= 0, \quad H\coloneqq \ybar - \int_0^{\thetabar} \frac{1}{1-\avg {B}\cos{2\psi}}\left(\avg {U_IB} \cos\psi\cos{2\psi} -\frac P{E^{12}}\sin{\psi}\right)\intd{\psi}.
\end{equation}
Thus, with $H$ constant, we have that $\ybar$ is an integral of a smooth, $2\pi$-periodic integrand, up to an additive constant. Furthermore, we have that
\begin{equation}
    \int_{0}^{2\pi} \frac{1}{1-\avg {B}\cos{2\psi}}\left(\avg {U_IB} \cos\psi\cos{2\psi} -\frac P{E^{12}}\sin{\psi}\right)\intd{\psi}=0
\end{equation}
by parity arguments and the periodicity of the integrand. Hence, $\ybar$ is bounded and periodic for all time. It inherits the period of $\thetabar$, which here is given by
\begin{equation}   
    \frac{2\pi}{q} = \frac{2\pi}{(a^2-c^2)^{1/2}} = \frac{2\pi}{\abs{E^{12}}(1-\avg {B}^2)^{1/2}}
\end{equation}
in units of the slow timescale.

Now suppose that $\abs{\avg{B}}\geq1$, a case that requires extreme shape elongation \citep{Bretherton1962}. This gives rise to fundamentally different dynamics, with the swimmer no longer tumbling. Instead, its angle asymptotes to a constant that, in turn, induces a slow drift to infinity along the $\e_2$ direction (perpendicular to the flow direction) for large time. This is even true for reciprocal swimmers if $\avg{U_IB}\neq0$. In other words, it is possible for a highly elongated reciprocal swimmer to self-propel indefinitely across pathlines in a stationary shear flow via the interaction between shear flow and the swimmer deformation.

\begin{figure}
 \centering
 \begin{overpic}[width=0.8\textwidth]{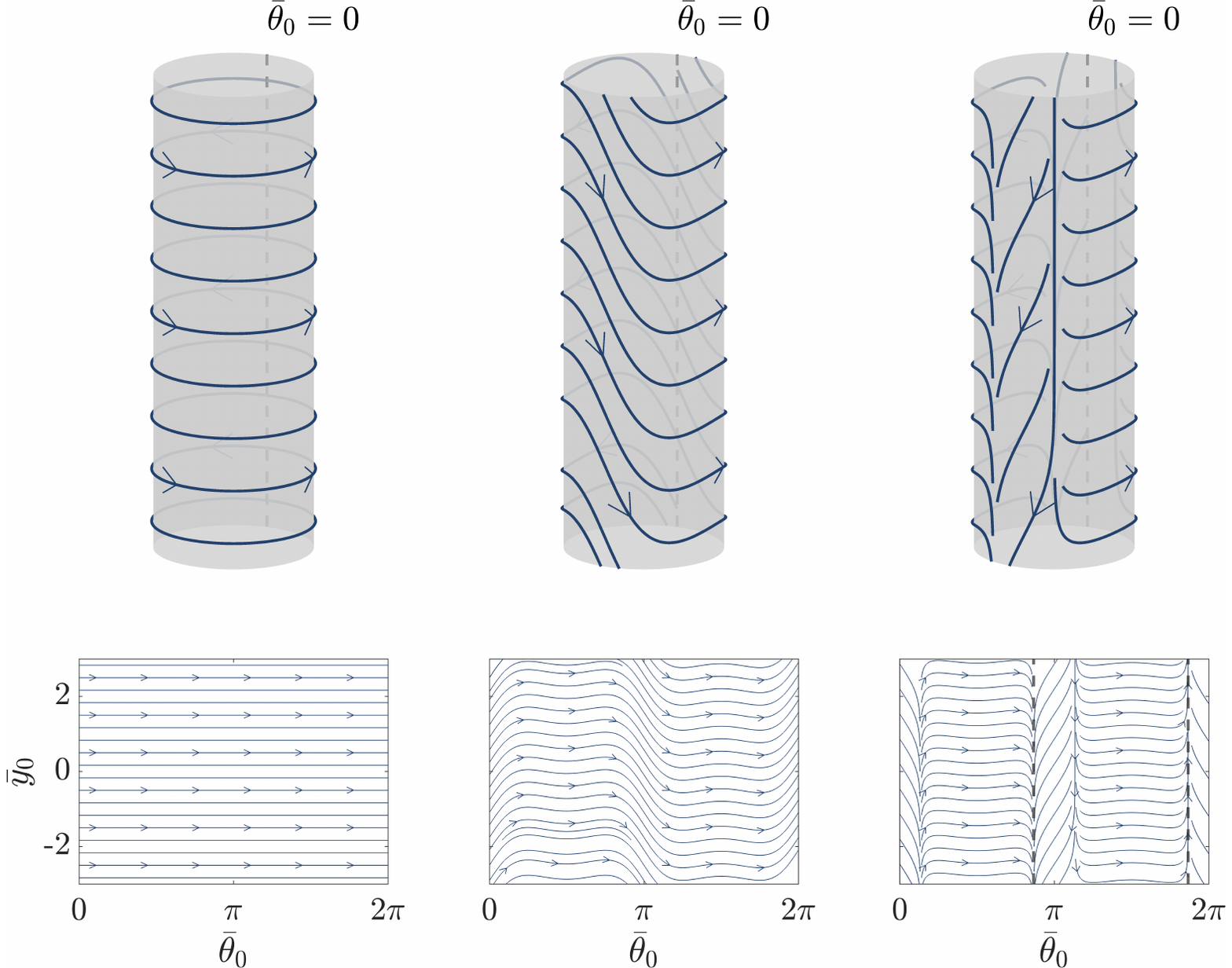}
     \put(2,68){(a)}
     \put(36,68){(b)}
     \put(70,68){(c)}
     \put(2,26){(d)}
     \put(36,26){(e)}
     \put(70,26){(f)}
 \end{overpic}
 \caption{Systematically averaged dynamics of a reciprocal swimmer in stationary shear flow. Shown both as dynamics on a cylinder and in the plane, we illustrate the semi-periodic phase space that corresponds to a reciprocal swimmer in stationary shear flow. We showcase three qualitatively distinct regimes: in (a,d), we have $(\avg{U_I B}, \avg{B}) = (0,0.5)$, leading to no motion at leading order; in (b,e), we have $(\avg{U_I B}, \avg{B}) = (1,0.5)$ and long-time periodic motion; in (c,f), we have $(\avg{U_I B}, \avg{B}) = (1,1.5)$ and progression. In (f), the dashed lines correspond to stable states of the angular dynamics.}
 \label{fig: stationary_shear_reciprocal}
\end{figure}

The range of possible leading order dynamics for a reciprocal swimmer ($P=0$) in various flows is illustrated in \cref{fig: stationary_shear_reciprocal}, with \cref{fig: stationary_shear_reciprocal}b,e showcasing periodicity on a long-timescale Jeffery's orbit. Non-trivial motility is highlighted in \cref{fig: stationary_shear_reciprocal}c,f, in which the reciprocal swimmer approaches a steady state of the angular dynamics and achieves net propulsion across pathlines of the flow.

\subsection{Motility in oscillatory shear}
A natural generalisation of stationary shear flow is oscillatory shear flow. For such a flow, we have $E^{11}=0$ and $\Omega^*(T)=-E^{12}(T)\neq0$, without loss of generality. The equations of motion simplify to
\begin{subequations}
\begin{align}
    \diff{\avg{x}_0}{t} &= 2\avg{E^{12}} \avg{y}_0 +  \avg{E^{12} U_I}\sin{\avg{\theta}_0} + \avg{E^{12} U_I B}\sin{\avg{\theta}_0}\cos{2\avg{\theta}_0}+P\cos{\avg{\theta}_0},  \\
    \diff{\ybar}{t} &= \avg{E^{12}\UI}\cos{\thetabar} - \avg{E^{12}\UI\B}\cos{\thetabar}\cos{2\thetabar} +P\sin{\avg{\theta}_0}, \\ 
    \diff{\thetabar}{t} &= - \avg{E^{12}} \left(1- \frac{\avg{E^{12}\B}}{\avg{E^{12}}} \cos{2\thetabar}\right),\label{eq: oscillatory shear: angle}
\end{align}
\end{subequations}
noting the appearance of an effective Bretherton parameter of $\avg{E^{12}\B}/\avg{E^{12}}$ in \cref{eq: oscillatory shear: angle}. Importantly, this quantity need not have magnitude less than unity, even if $\abs{B(T)}<1$ for all $T$. 

Below, we only consider cases where the effective Bretherton parameter has a magnitude greater than unity, noting that this does not necessitate the geometrical constraint of severe elongation, though it does mean the angular dynamics is asymptoting. In this case, the angular dynamics evolves to an asymptotically constant angle at large time. This holds even if the average shear rate is zero (i.e. $\avg{E^{12}} =0$) but $\avg{E^{12}\B}\neq 0$, in which case we instead have 
\begin{equation}
   \diff{\thetabar}{t} = \avg{E^{12}\B}\cos{2\thetabar} 
\end{equation}
and the angular dynamics evolves to an asymptotic state with $\cos{2\thetabar}=0$.

If, for the moment, we assume that $\avg{E^{12}}\neq 0$ and define $\mathcal{C} = \avg{E^{12}}/\avg{E^{12} B} \in [-1,1]$ as the reciprocal of the effective Bretherton parameter, the long-time asymptote for $\cos{2\thetabar}$. Then define
\begin{subequations}\label{Acal}
\begin{align}
    \mathcal{A} &\coloneqq \left. (\avg{E^{12} U_I}\sin{\avg{\theta}_0} + \avg{E^{12} U_I B}\sin{\avg{\theta}_0}\cos{2\avg{\theta}_0}+P\cos{\avg{\theta}_0})\right\rvert_{\cos2\thetabar=\mathcal{C}}\\ 
    &= \pm\frac{\sqrt{1-\mathcal{C}}}{\sqrt{2}} 
    \left( \avg{E^{12} U_I}+\avg{E^{12} U_I B}\mathcal{C} \right) + \frac{\sigma P\sqrt{1+\mathcal{C}}}{\sqrt{2}} \\ 
    \mathcal{B} &\coloneqq \left. (\avg{E^{12}\UI}\cos{\thetabar} - \avg{E^{12}\UI\B}\cos{\thetabar}\cos{2\thetabar}+P\sin{\avg{\theta}_0})\right\rvert_{\cos2\thetabar=\mathcal{C}} \\ 
    &= \frac{\sigma\sqrt{1+\mathcal{C}}}{\sqrt{2}}\left(\avg{E^{12}\UI}-\avg{E^{12}\UI\B} \mathcal{C}\right) \pm\frac{P\sqrt{1-\mathcal{C}}}{\sqrt{2}},
\end{align} 
\end{subequations}
where $\sigma \in\{-1,1\}$ and the sign choices ultimately depend on initial conditions of the dynamics and the sign of $\mathcal{C}$. With this simplifying notation, for large time we have
\begin{equation}
   \diff{\avg{y}_0}{\avg{x}_0} \approx \frac{\mathcal{B}}{\mathcal{A}+2\avg{E^{12}} \avg{y}_0 }. 
\end{equation}
Neglecting the asymptotically small errors in this approximation, one may integrate to give 
\begin{equation} \label{Kcal} \mathcal{B}  \avg{x}_0 = \mathcal{K}+ \mathcal{A} \avg{y}_0+\avg{E^{12}} \avg{y}_0^2, 
\end{equation}
where $\mathcal{K}$ is a constant. Thus, excluding possible edge cases, the long-time swimmer trajectories are parabolic, regardless of the details of the flow and the swimming. One such edge case is $\mathcal{B}=0$, which gives the limiting case of a line, for instance.

\begin{figure}
 \centering
 \begin{overpic}[width=0.8\textwidth]{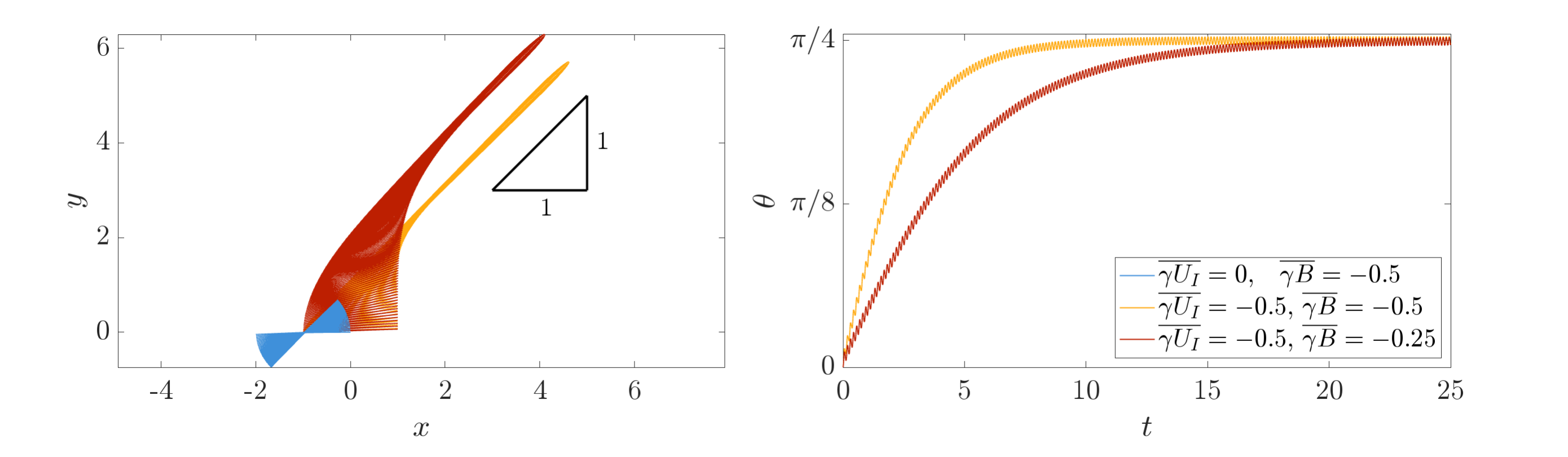}
    \put(3,28){(a)}
    \put(48,28){(b)}
 \end{overpic}
 \caption{Behaviours of reciprocal swimmers in unsteady shear flow with zero average shear rate. Note in (a) that the long-time trajectory has a gradient of magnitude unity, a universal prediction for swimmers with sufficient symmetry (such as maintaining a shape that is always a body of revolution with fore-aft symmetry).}
 \label{fig: unsteady shear reciprocal}
\end{figure}

Finally, for the case where there is no net shear, we have $ \avg{E^{12}}=0$ and assume that $\avg{E^{12}B} \neq0$. Here, the trajectory is also linear, even though naive averaging would predict that the reciprocal swimmer has no net motion. In particular, we have $\mathcal{C}=0$, whereby explicit calculation reveals extensive simplification with $\mathcal{A}/\mathcal{B}=\pm 1$ in the large time limit, regardless of whether $P=0$ or $P\neq 0$. Hence, we have
\begin{equation}
    \diff{\avg{y}_0}{\avg{x}_0} \approx \pm1,
\end{equation}
so that the trajectories are simply straight lines with gradient $\pm$1, independent of the details of the flow, swimmer, and initial conditions except for the choice of sign, at least once the angular dynamics is asymptoting. Example such trajectories are shown for a reciprocal swimmer in \cref{fig: unsteady shear reciprocal}a, with the associated angular dynamics illustrated in \cref{fig: unsteady shear reciprocal}b. This is an explicit example of a reciprocal swimmer in a zero-mean oscillating shear swimming across pathlines, breaking Purcell's scallop theorem. Notably, the swimmers in this example may be highly symmetric, and extensive swimmer elongation is not required to be in this dynamical regime.

\section{Discussion and conclusions}\label{sec: discussion}

We have considered swimmers that are characterised by a separation of timescales, with motility driven by fast-timescale processes such as swimmer treadmilling and shape changes. These drivers are independent of the background flow and induce motility associated with a slower timescale, as frequently observed \citep{Smith2009a,Curtis2013,Ishimoto2014a,Pak2015}. The swimmer may also induce its own rotation on the fast timescale, though we have assumed a degree of symmetry throughout the motion. In particular, the swimmer is assumed to possess helicoidal symmetry together with additional symmetries associated with reflection planes or rotation axes. These symmetries are summarised in \cref{geshs}, with the simplest case corresponding to a body of revolution and fore-aft symmetry.

Within this framework, we have derived the equations of motion for both the swimmer translational and angular motility, making use of multiscale asymptotic methods that exploit the ratio of timescales  to generate simplified equations for the leading order dynamics on the slow timescale. As expected given the restrictions imposed on the swimmer, any motion perpendicular to the plane of the flow decouples and we essentially neglect this trivial aspect of the flow. The angular dynamics and the planar translation dynamics are given by a rapid periodic oscillation induced by the swimmer activity, superimposed with angular changes and trajectories that evolve on the slow timescale.

The full system of governing equations is intricate, involving the large number of variables and parameters summarised in \cref{Tab1,Tab1.1}. In analysing the slow dynamics, however, it is clear that the resulting motion can be extensively characterised by only two groups of variables. These are
\begin{equation}\label{twoparams} 
    \frac{a^2}{b^2+c^2}\quad \text{ and } \quad \frac{\avg{\Omega^*}^2}{\avg{E^{11}}^2+\avg{E^{12}}^2},
\end{equation}
which describe the background flow and the swimming activity in turn. Here, $a$ is the fast-timescale average of the angular velocity of the swimmer and the background flow, while $b^2+c^2$ measures the impact of the fast-timescale average of the fluid rate of strain on the swimmer's angular dynamics. Similarly, 
$\avg{\Omega^*}$ is the fluid angular velocity and $(\avg{E^{11}}^2+\avg{E^{12}}^2)^{1/2}$ is a measure of the rate of strain, both averaged over the fast timescale. Hence, the behaviour of the system is characterised by whether or not angular velocity dominates the impact of rate of strain for both the swimmer and the fluid. 

In particular, whether the swimmer endlessly tumbles or instead asymptotes to a fixed angle of swimming depends only on $a^2/(b^2+c^2)$, which can in turn depend strongly on the level of swimmer symmetry and the background flow. Hence, determining the angular dynamics of a swimmer involves detailed knowledge of both the swimmer and the background flow. In contrast, the character of the translational motion depends only on the flow and not on the properties of the swimmer, splitting into trajectories that are exponential, oscillatory or linear in time based on the value of $\avg{\Omega^*}^2/(\avg{E^{11}}^2+\avg{E^{12}}^2)$. In combination, these observations of angular dynamics and motility can extensively inform qualitative features of the swimmer trajectory. This ability to extensively classify the behaviours of a swimmer via only a two-dimensional parameter space is much simpler than one might initially anticipate. We present a summary of this classification in \cref{Tab2}.

\setlength\extrarowheight{3pt}
\begin{table}
\smaller 
\begin{NiceTabular}{|C{1.9cm}|C{2.75cm}|C{8.15cm}|}
\hline 
Angular dynamics & Translational dynamics & Trajectories and observations    \\ \Hline\Hline  
 Asymptoting  or tumbling & Exponential $\avg{E^{11}}^2+\avg{E^{12}}^2>\avg{\Omega^*}^2 $ & The trajectory drifts to infinity at an exponential rate in the slow timescale once the average background flow strain rate dominates the average angular velocity, independent of the swimmer details.  \\  \hline 
Asymptoting ~~$a^2\leq b^2+c^2 $ & Oscillating $ \avg{E^{11}}^2+\avg{E^{12}}^2<\avg{\Omega^*}^2 $ & The long-time swimmer trajectory is an oscillation on the slow timescale of period $2\pi/(\avg{\Omega^*}^2-\avg{E^{11}}^2-\avg{E^{12}}^2)^{1/2}$. Progressive swimmer motion is converted to oscillatory motion by the background flow, independent of the swimmer details. \\  \hline 
Tumbling $a^2> b^2+c^2 $ & Oscillating $ \avg{E^{11}}^2+\avg{E^{12}}^2<\avg{\Omega^*}^2 $ & Whether the long-time dynamics is oscillatory or unbounded depends on whether resonance occurs. Resonance can require parameter fine-tuning, but not always, e.g. a body of revolution with fore-aft symmetry in a rotational background flow. \\  \hline 
Asymptoting $a^2\leq b^2+c^2 $ & Linear, $\mat{A}= \vec{0}$, $ \avg{E^{11}}^2+\avg{E^{12}}^2=\avg{\Omega^*}^2 $ & Trivial background flow is excluded. The swimmer will drift to infinity linearly in time if the net swimming speed $P$ is non-zero. Infinite drift can also occur even if there is only reciprocal swimming, highlighting  that Purcell's theorem can be broken by a zero-mean oscillatory flow. \\  \hline 
Tumbling $a^2> b^2+c^2 $ & Linear, $\mat{A} =\vec{0}$, $ \avg{E^{11}}^2+\avg{E^{12}}^2=\avg{\Omega^*}^2 $ & The swimmer does not drift to infinity, independent of the swimmer details, except fore-aft asymmetry is necessary to satisfy the tumbling  condition. \\  \hline 
Asymptoting $a^2\leq b^2+c^2 $ & Linear, $\mat{A}\neq \vec{0}$, $ \mat{A}^2=\vec{0}$, $ \avg{E^{11}}^2+\avg{E^{12}}^2=\avg{\Omega^*}^2 $ & These flows are equivalent to a shear flow. The swimmer will drift to infinity, with even reciprocal swimming capable of generating motion perpendicular to the pathlines, breaking Purcell's theorem. \\  \hline 
Tumbling $a^2> b^2+c^2 $ & Linear, $\mat{A}\neq \vec{0}$, $\mat{A}^2=\vec{0}$, $ \avg{E^{11}}^2+\avg{E^{12}}^2=\avg{\Omega^*}^2 $ & These flows are equivalent to a shear flow. The swimmer will drift to infinity along the pathlines but will not drift indefinitely perpendicular to the pathlines, independent of the swimmer details. \\  \hline 
\end{NiceTabular}
 \begin{caption}{\label{Tab2} A summary of swimmer behaviours in planar linear background flows. Edge cases, where parameter fine tuning leads to behaviours distinct from the more general cases, are not summarised here.}  
 \end{caption}
\end{table}

One can also immediately note instances where the interaction of the swimmer with the background flow can induce progressive motion even for reciprocal swimming, where no net motion is generated in the absence of background flow.   This occurs with motion across pathlines for shear flows and  mean-zero oscillatory flows  if the swimmer acquires a fixed angle at long time, as further illustrated for the special cases with an oscillating shear flow. Thus, the interaction of the swimmer with a background flow provides another means to breaks Purcell's scallop theorem that supplements other mechanisms, such as the introduction of viscoelasticity or inertia \citep{lauga2007a,lauga2011,qiu2014,derr2022}. 
Conversely, the characterisation of the swimmer also highlights when progressive swimming can be converted to oscillatory trajectories by the background flow, in particular for oscillatory flows where the swimmer angle asymptotes to a constant for large times. 

As well as these general considerations, special cases of the dynamics for these swimmers in planar flows were considered, restricting attention to specific background flows and highly symmetrical swimmers. Our first example considered a rotational flow, where tumbling is observed for the symmetric swimmer considered. In turn, this induced resonance once the swimmer had a non-zero net swimming speed $(P\neq 0)$. This also demonstrates that swimmer motility need not be converted to oscillations for sufficiently symmetric swimmers in rotating flows. Nonetheless, it is possible for the swimmer to enter the regime of an asymptoting angle despite the presence of a rotational background flow. For example, it may break fore-aft symmetry to rotate in the opposite direction to the background flow angular velocity. Then, in this regime, progressive motility is converted to oscillation (e.g. \cref{Tab2}). Hence, we can observe that, despite numerous observations in \cref{Tab2} being independent of the details of the swimmer, aspects of inertialess motility in background flows can be sensitive to the details of the swimming gait. For instance, the gait may allow passage between the different types of behaviour in parameter space associated with the parameter groupings of \cref{twoparams}. 

For irrotational flows with a highly symmetric swimmer, we observe that the swimmer has an asymptoting angle for large time and the long-time trajectory forms a hyperbola in the plane of the flow. Furthermore, the observation of hyperbolic trajectories is robust to many features of the swimmer, though this can be disrupted by the introduction of a sufficiently large average angular velocity, $\Omega(T)$, induced by the swimmer in the absence of a background flow, which requires a fore-aft symmetry breaking of the swimmer. 

In shear flows, there is in general a drift in the direction of fluid flow, but motility across pathlines is also possible. For stationary shear and a highly symmetric swimmer, this can only occur with extensive swimmer elongation, as such dynamics requires that the Bretherton parameter $B$ satisfy $B^2>1$. Otherwise, the dynamics across pathlines is oscillatory. In contrast, for background shear flows that have an oscillating contribution, such extreme elongation need not be required for indefinite drift across pathlines. More generally, for highly symmetric swimmers that do not tumble, the long-time trajectory is in the shape of a parabola. Furthermore, if the mean shear flow is zero for a highly symmetric non-tumbling swimmer, the trajectory reduces to a line with a gradient that has modulus unity (given one axis aligned along the direction of the flow). The latter is one of many examples of the previously known observation that a priori averaging, that is the averaging of oscillations without considering the details of the particular model, can generate incorrect results, as reported for example by \citet{Walker2021Yawing,Walker2023}.

While swimmers in shear flow have been subject to extensive study, for instance by \citet{KarpBoss2000,Hope2016b,Gaffney2022b} and \citet{Walker2022}, we see that swimmer behaviour in a pure steady shear flow truly is a special case. The introduction of oscillations can fundamentally change the character of swimmer behaviour, leading to parabolic trajectories emerging. Similarly, small changes to the background flow can extensively change the swimmer behaviour. For example, a small change in the flow so that the flow angular velocity dominates the rate of strain (if only weakly) induces an oscillatory motion of a non-tumbling swimmer rather than a drifting motion. In contrast, if the small change is such that the rate of strain dominates then the drift to infinity switches to become exponential in time, as may be inferred from \cref{Tab2}.

This raises the question of how further changes in the flow influence swimmer dynamics. The consideration of non-linear flows and spatially non-constant flow angular velocities and rates of strain are pertinent examples left for future work. A further question concerns rheotaxis, which has been observed and predicted for swimmers in Poiseuille flow \citep{Omori2022,Walker2022er} and sperm cells under relatively general circumstances, for instance swimming in shear flows \citep{Miki2013,Kantsler2014,Ishimoto2015}, as well as predicted for squirmers close to a no-slip wall \citep{Uspal2015,Ishimoto2017}. 
These observations, however, do not fall into the remit of the analysis presented here in that they involve a reorientation in the swimming plane, perpendicular to $\e_3=\hat{\e}_3$, due to a flow that \emph{varies} in the $\e_3$ direction, while we have only considered flows that are in the swimming plane and not perpendicular to it, except for the trivial case of a constant flow in this direction. Hence, such observations do not contradict the observation here that, at leading order in $1/\omega$, no predictions of rheotaxis have emerged for a swimmer in a plane in response to a linear flow restricted to the same plane. The absence of rheotaxis is  illustrated by the hyperbola and the parabola of \cref{hyp,Kcal} in special cases, for example, and is also apparent from the prediction that the final swimming direction angle for an asymptoting swimmer depends on the initial swimming direction, as may be inferred from \cref{tsol}.

Since individual dynamics often feature in the construction of population models for swimmers \citep{Saintillan2013,Ezhilan2013,Junot2019}, this study also offers the prospect of facilitating the development of collective swimmer models in more general background flows than has typically been considered. Furthermore, given the classification of swimmer behaviours, one can also consider how one may control a swimmer (or mobile microrobot) in a background flow, assuming that it cannot swim with sufficient speed to render the background flow as a small perturbation that can be ignored in terms of navigation. In particular, manipulating the swimmer cannot control the trichotomy of the translational dynamics (into those of exponential, oscillating and linear character) as these depend only on the background flow. However, the swimmer can always, at least in principle, be switched from tumbling to asymptoting in its angular behaviour, for instance by controlling its self-induced rotation $\Omega(T)$. While such switching does not have an impact on the exponential translation case, switching to tumbling angular dynamics in the case of linear translation increases the prospect of localised trajectories rather than drifting, with potentially the opposite for oscillating flows if a resonance occurs in the latter case with tumbling, as again may be inferred from  \cref{Tab2}.

In summary, the equations of motion for a swimmer possessing modest spatial symmetry in a linear planar background flow have been derived using the assumption of inefficiency. That is, we have assumed that the swimmer's net motion is much slower than the deformations and treadmilling required to generate its motion, noting that this is commonplace in microswimming. The resulting solutions allow for a classification of swimmer behaviour based on the ratio of the background flow angular velocity to rate of strain, and the ratio of the swimmer and flow angular velocity to the swimmer's interaction with the flow rate of strain. Thus, a complex system characterised by numerous variables can be extensively understood in terms of only two degrees of freedom. Furthermore, the presented study further highlights the need for careful averaging in analysing the equations of motion, whilst observing that the interactions between a swimmer and a background flow can provide a further mechanism for circumventing Purcell's scallop theorem. Common, nearly universal behaviours are also predicted, such as long-time parabolic trajectories for swimmers in oscillatory shear flows. Furthermore, the examples considered here highlight when swimmer navigation in background flows is futile, together with when and how the swimmer is capable of switching from localised trajectories to inexorable drift, or vice-versa, enabling an element of rational control over swimmer and microrobot movement in linear background flows. 

{\bf Acknowledgements}
K.I. acknowledges the Japan Society for the Promotion of Science (JSPS) KAKENHI (Grant No. 21H05309) and the Japan Science and Technology Agency (JST), FOREST (Grant No. JPMJFR212N). B.J.W. is supported by the Royal Commission for the Exhibition of 1851.

{\bf Data and rights retention statement}
All codes used to generate figures in this manuscript are available at \url{https://github.com/Mar5bar/multi-timescale-microswimmers-in-background-flows}. For the purpose of Open Access, the authors will apply a CC BY public copyright licence to any Author Accepted Manuscript (AAM) version arising from this submission.

Declaration of Interests. The authors report no conflict of interest.
 
\appendix
\section{Deriving the governing equations}\label{appendix1}
Here, we derive the governing equations for the translation and rotation of shape changing and treadmilling swimmers in time-dependent planar linear background flows, assuming that the swimmer moves  in the plane of the flow and with sufficient swimmer symmetry, as detailed in \cref{sgedgc}, to ensure a relatively simple generalisation of idealised models. We first reduce the problem to that of treating the swimmer at a fixed time as a rigid particle, enabling rigid particle methods, such as that presented in the Appendices of \citet{Dalwadi2024b}.

We inherit the notation of the main text, for example with $\x = x \e_1+y\e_2+z\e_3$ denoting the position of a point relative to the laboratory-frame basis, $\{\e_1, \e_2, \e_3\}$, while $\{\ehat{1}, \ehat{2}, \ehat{3}\}$ denotes the swimmer frame basis, with origin at $\x_c$, the swimmer centroid. Hence, 
\begin{equation} \label{ct1} 
\ehat{1}=\cos\theta \e_1 +\sin\theta \e_2,
 \quad\ehat{2}=-\sin\theta \e_1 +\cos\theta \e_2,\quad\ehat{3}=\e_3, 
\end{equation} 
and the background flow field is given by 
\begin{subequations}
\begin{align}
\flowVel(\x,T) &= \flowVel_{tr}(T) + \flowAngVel(T)\wedge \x + \flowStrainRate(T)\x \\ 
&= \flowVel_c + \flowAngVel(T)\wedge (\x-\x_c) + \flowStrainRate(T)(\x-\x_c), \label{exp0app}
\end{align} 
\end{subequations}
where $T=\omega t$ is the fast timescale, $ \flowVel_{tr}(T) $ denotes the spatially constant translational aspect of the background flow. Here,  $\flowStrainRate (T)$ and $\flowAngVel (T)$ denote the rate of strain and angular velocity, respectively, which are spatially constant by flow linearity. Below, we use $\flowVel_c = \flowVel(\x_c(t,T),T)$ to denote the background flow at the swimmer centroid if the swimmer was absent from the domain.
 
\subsection{Model mechanics and the Grand Mobility Tensor}
With $\vec{u}$ denoting the velocity vector field of the flow \emph{including} the swimmer, in contrast to the background flow $\flowVel$ which \emph{excludes} the swimmer, we have that the fundamental equations for the displacement flow, $\vec{u}^{d_p}=\vec{u} -\flowVel$, with disturbance pressure $p^{d_p}$ are given by 
\begin{equation} \label{se} 
 \nabla p ^{d_p} = \nabla^2\vec{u}^{d_p}, \quad \nabla \cdot \vec{u}^{d_p} =0,
\end{equation} 
exterior to the particle with decay boundary conditions at spatial infinity. Noting that the swimmer shape deformations and treadmilling are independent of the external flow, the boundary conditions for $\vec{u}$ on the swimmer surface, $\x \in \partial \Lambda$, are given by 
\begin{subequations}
\begin{align}
\vec{u}^{d_p} (\x) &= \vec{u}^{S}(\x, T) + \vec{U}(T,\flowVel) + \vec{\Omega} (T,\flowVel) \wedge(\x-\x_c)- \flowVel(\x,T), \\ 
&=\vec{u}^{S}(\x, T) + \left[\vec{U}(T,\flowVel) -\flowVel_c \right]\nonumber\\
&\qquad+(\vec{\Omega} (T,\flowVel) -\flowAngVel (T) )\wedge(\x-\x_c) - \flowStrainRate(T)(\x-\x_c).
\label{bc} 
\end{align}
\end{subequations}
Here, $\x_c=\x_c(t,T)$ is the location of the swimmer centroid and $\flowVel_c=\flowVel_c(\x_c(t,T),T)$ is the background flow at the swimmer centroid. In addition, $\vec{u}^{S}(\x,T)$ is the surface velocity, accommodating shape-shifting and tread-milling, which is $2\pi$-periodic in $T$, while $\vec{U}(T,\flowVel)$ denotes the swimming speed of the particle in the background flow, $
\flowVel$, at time $T$. Here, $\vec{\Omega}(T,\flowVel) $ denotes the angular velocity of the body fixed frame relative to the inertial frame in the background flow, $
\flowVel(\x,T)$.

As the linear velocity and angular velocity are six degrees of freedom in the unknowns, six further constraints are required; assuming the swimmer is not also being driven by an external forcing, such as a magnetic field, these are no net force and torque, that is 
\begin{equation} \label{bcft} 
\int_{\partial \Lambda} \vec{\sigma} \cdot \vec{n} \mathrm{d}S = \int_{\partial \Lambda} (\x-\x_c)\wedge \vec{\sigma} \cdot \vec{n} \mathrm{d}S = \vec{0}, 
\end{equation} 
where $\vec{n}$ is the normal, defined to point \emph{out} of the fluid domain, and 
\begin{equation}
    \vec{\sigma} = - p^{d_p} \vec{I} + (\nabla \vec{u}^{d_p} + (\nabla \vec{u}^{d_p})\transpose )
\end{equation}
is the Cauchy stress. A pressure gauge condition is also required to pin the translational freedom in the pressure $p^{d_p}\mapsto p^{d_p} + \text{constant}$. 

We decompose the disturbance problem, \crefrange{se}{bcft},
into two auxiliary problems for the pressure and 
velocity fields $(p^{d_1},\vec{u}^{d_1})$ and $(p^{d_2},\vec{u}^{d_2})$, respectively. 
The first auxiliary problem is for the swimmer shape at fixed time $T$, with the shape pinned within quiescent fluid without the freedom to translate or rotate, but nonetheless undergoing the shape shifting and treadmilling surface changes. Thus, the bulk equations, \cref{se}, are inherited as are the decay boundary conditions at spatial infinity, but the velocity boundary condition becomes 
\begin{equation}
    \vec{u}^{d_1} = \vec{u}^{S}(\x, T) , \quad \x \in \partial \Lambda.
\end{equation}
Given a pressure gauge fixing, no further force or torque constraints are required, as the constraints are now the absence of translation and rotation. Let $ \vec{F}(T)$ define the force required to be imposed on the particle (for example by micro-tweezers in practice) to enforce these constraints, and analogously for $\vec{T}(T)$. The net force and torque on the particle must be zero in the inertialess limit and, hence,
\begin{equation} \label{netft} 
\int_{\partial \Lambda} \vec{\sigma}^{d_1} \cdot \vec{n} \intd{S} + \vec{F}(T) = \vec{0}, \quad \int_{\partial \Lambda} (\x-\x_c)\wedge \vec{\sigma}^{d_1} \cdot \vec{n} \intd{S} + \vec{T}(T)= \vec{0} . 
\end{equation} 

The second auxiliary problem similarly inherits the bulk Stokes equations, \cref{se}, the decay boundary conditions at spatial infinity and the pressure gauge condition. However, with fixed $t,T$, the velocity boundary condition is taken to be 
 \begin{equation}
 \vec{u}^{d_2} (\x)= \left[\vec{U}(T,\flowVel) -\flowVel_c \right] + 
\left[\vec{\Omega} (T,\flowVel) -\flowAngVel(T)\right]\wedge(\x-\x_c) - \flowStrainRate(T)(\x-\x_c),~~~\label{bcud2}
\end{equation} 
where $\vec{U}(T,\flowVel)$ is the a priori unknown translation speed of the particle in the background flow at this instant for this problem, and $\vec{\Omega}(T,\flowVel) $ is the a priori unknown angular velocity of the swimmer frame relative to the inertial frame in the background flow for this problem. The six additional constraints are taken to be 
\begin{equation} \label{bcftnew} 
\int_{\partial \Lambda} \vec{\sigma}^{d_2} \cdot \vec{n} \intd{S} = \vec{F}(T), \quad \int_{\partial \Lambda} (\x-\x_c)\wedge \vec{\sigma}^{d_2} \cdot \vec{n} \intd{S} = \vec{T}(t), 
\end{equation} 
so that $(p^{d_1}+p^{d_2},\vec{u}^{d_1}+\vec{u}^{d_2})$ provide a solution of the original problem, which is unique given pressure gauge fixing and, thus, this is \emph{the} solution. Hence, $\vec{U}(T,\flowVel)$ and  $\vec{\Omega} (T,\flowVel)$ in the solution of the above second auxiliary problem are also the swimming speed and angular velocity of the shape shifting and tread-milling swimmer in the background flow. 

Note that, in the second auxiliary problem, the velocity boundary condition is that of a rigid particle, with an unknown net force and an unknown net torque prescribed, which can be framed in the setting of the Grand Mobility Tensor framework of \citet{Kim2005}. In particular, we proceed by first considering the behaviour of the swimmer in a quiescent fluid, before then investigating the swimmer dynamics in the background flow, under the assumption that the shape-shifting and tread-milling are unchanged in the presence or absence of the background flow. 

In the absence of a background flow, $\flowVel_c=\flowAngVel =\vec{0}$,  $\flowStrainRate=\vec{0}$ and the surface velocity $\vec{u}^{S}(T)$ induce a swimming speed $\vec{U}(T,\vec{0})$ and angular velocity $\vec{\Omega}(T,\vec{0})$. In the framework of the Grand Mobility Tensor of \citet{Kim2005} for the second auxiliary problem, with a trivial background flow, we have at time $T$ that 
 \begin{equation} \label{gmt1} 
 \begin{pmatrix}
 -\vec{U}(T,\vec{0}) \\ 
 -\vec{\Omega}(T,\vec{0})\\ 
 \vec{S} 
 \end{pmatrix} =
 \begin{pmatrix}
\mat{A} (T)& \tilde{\tensor{b}}(T) &\tilde{\tensor{g}}(T)\\
\tensor{b} (T)& \tensor{c}(T) & \tilde{\tensor{h}}(T) \\
\tensor{g} (T)& \tensor{h}(T) & \tensor{m}(T) \\
 \end{pmatrix}
 \begin{pmatrix}
\vec{F} (T) \\ 
 \vec{T}(T) \\ 
 \vec{0}
 \end{pmatrix}, 
 \end{equation} 
where the $T$-dependent block entries of the grand mobility tensor relate the force, $\vec{F}(T)$, to the velocity of the particle in a quiescent field. While not used here, $\vec{S}$ denotes the stresslet associated with the particle motion. Reinstating the background flow with the shape shifting and treadmilling assumed unchanged, so that $\vec{F} (T)$, $\vec{T}(T)$ and the Grand Mobility Tensor associated with the second auxiliary problem are invariant with the change of external flow, the analogous \citet{Kim2005} relation with a background flow is given by 
\begin{equation} \label{gmt2} 
 \left( \begin{array}{c}
\flowVel_c -\vec{U}(T,\flowVel) \\ 
 \flowAngVel_c - \vec{\Omega}(T,\flowVel) \\ 
 \vec{S}^{*}
 \end{array}
 \right) =
 \begin{pmatrix}
\tensor{A} (T)& \tilde{\tensor{b}}(T) &\tilde{\tensor{g}}(T) \\
\tensor{b} (T)& \tensor{c}(T) & \tilde{\tensor{h}}(T) \\
\tensor{g}(T)& \tensor{h}(T) & \tensor{m}(T) \\
 \end{pmatrix} 
 \begin{pmatrix}
\vec{F} (T) \\ 
 \vec{T}(T) \\ 
 \flowStrainRate 
 \end{pmatrix}, 
\end{equation}
where $\vec{S}^{*}$ denotes the stresslet associated with the particle motion in the planar, uni-directional background flow. 

\subsection{The swimmer velocity and the governing equations}\label{svage}
The objective is to use \cref{gmt1,gmt2} to determine $\vec{U}(T,\flowVel)$ and $\vec{\Omega}(T,\flowVel)$, thus yielding the equation of motion. This requires us to specify $\vec{U}(T,\vec{0})$, the linear velocity of the swimmer in the absence of flow, and $\vec{\Omega}(T,\vec{0})$, the angular velocity of the swimmer in the absence of flow. The linear velocity scales with the frequency of the shape deformations and treadmilling by the linearity of Stokes flow and the fact that time is simply a parameter in the absence of temporal derivatives. Thus, we can write the velocity of the swimmer for $\flowVel=0$ in the form 
\begin{equation}\label{veleq} 
\vec{U}(T,0)=\omega U(T)\ehat{1} + \omega P^* \ehat{1}, 
\end{equation}
where $\omega\gg1$ is the scale of the swimmer deformation speed relative to the background flow speed, $\omega P^*$ is the average speed along the body axis and $U(T)$ is the oscillatory speed. The latter  averages to zero over a period, taken to be $2\pi$ without loss of generality. While $ \omega\abs{P^*}\sim \omega\sup_T \abs{U(T)}$ has been considered in previous analytically-based multiple timescale studies of swimmers in background flow \citep{Walker2022er}, many swimmers are inefficient and have small net swimming speeds compared to the velocity of oscillatory motions, or even a zero net-swimming speed in the case of reciprocal swimmers. This separation of scales is observed for many theoretical swimmers \citep{Curtis2013,Ishimoto2014a,Pak2015}, with the three-link swimmer an extreme example \citep{Curtis2013}, while it also observed for biological microswimmers. For example, considering the experimental observations of sperm in \citet{Smith2009a}, and restricting attention to the effectively Newtonian, low-viscosity medium, the progressive velocity of the cell is \SI{62}{\micro\metre\per\second}, while its flagellar wavespeed is \SI{920}{\micro\metre\per\second}. 

Hence, we consider scales where
\begin{equation}
    \omega\abs{P^*}\ll \omega\sup_T\abs{U(T)}.
\end{equation}
For many cases, $\omega \abs{P^*}\ll 1 $ is not of interest, as the swimmer will be washed out by the background flow, noting the latter has an $\ord{1}$ velocity scale by the non-dimensionalisation. However, for analysing whether Purcell's scallop theorem generalises to include swimmer-flow interactions specifically concerns reciprocal swimmers 
with $P^*=0$, so this case is also investigated in the main text. We also impose the restriction that $ \omega\abs{P^*}\not\gg 1$, so that there is a substantive interaction between the swimmer and the flow, rather than the swimmer being only weakly perturbed by the flow. Hence, we have 
\begin{equation}\label{UT0} 
\vec{U}(T,0) =\omega U(T)\ehat{1} + P \ehat{1} , \quad \abs{P} \sim \bigO{\sup_T \abs{U(T)}} \sim \bigO{1}.
\end{equation}
We additionally note that any corrections at higher powers of $1/\omega \ll 1 $ will not feature in the equations once the multiple scales approximation has been taken. Then, with use of \cref{gmt1} to eliminate $\vec{F}(T)$ and $\vec{T}(T)$, we have from \cref{gmt2} that the governing equations for the particle velocity and angular velocity in the shear flow are given by 
 \begin{equation} \label{gvt1} 
 \diff{\x_c}{t}=\vec{U}(T,\flowVel) = \flowVel_c + \omega U(T) \ehat{1}+ P\ehat{1} -\tilde{\tensor{g}}\flowStrainRate .
 \end{equation} 
 
Considering the angular dynamics, the restriction to planar dynamics ensures that we can write
\begin{equation}
    \vec{\Omega}(T,\flowVel) = \dot \theta \e_3 =\dot \theta \ehat{3}, \quad \flowAngVel(T)=\Omega^*(T)\ehat{3}, \quad  
 \vec\Omega(T,\vec{0})=\Omega(T,\vec{0})\ehat{3}.
\end{equation}
The term $\Omega(T,\vec{0})$ represents the angular velocity induced by the shape deformation when the swimmer is in background flow. As these deformations are on the fast timescale, we can write 
\begin{equation}
    \Omega(T,\vec{0}) = \omega\Omega_f(T) + \Omega(T),
\end{equation}
where $\Omega_f(T)$ is the angular velocity on the fast timescale and $\Omega(T)$ is the first correction in the expansion of $\Omega(T,\vec{0})$ in powers of $1/\omega \ll 1$. In particular, we maintain generality by including the possibility of a contribution at 
$\ord{1}$, analogously to the inefficient contribution to translation motion. Similarly, contributions at higher orders of $1/\omega\ll 1$ do not contribute once the multiple scales approximation has been imposed. 

However, in contrast to the translational equations, with $U(T)$ averaging to zero over a fast period, we do not rule out the possibility that swimmers may turn  efficiently, on the same timescale as the shape deformation. This may be immediately inferred, for example, from studies of sea urchin sperm chemotaxis such as \cite{Wood2005}, so that we do not assume that  $\Omega_f(T)$ averages to zero over a fast timescale period. Nonetheless, noting the memoryless property of Stokes flow, both $\Omega_f(T),\Omega(T)$ inherit the periodicity of the shape deformation and, thus, are both $2\pi$-periodic.
Then, eliminating $\vec{F}(T)$ and $\vec{T}(T)$ analogously to the derivation of \cref{gvt1}, we have 
\begin{subequations}
    \begin{align}
        \dot \theta \e_3 &= \dot \theta \ehat{3} = \vec{\Omega}(T,\flowVel) = \flowAngVel(T)+\vec{\Omega}(T,\vec{0}) - \tilde{\tensor{h}} \flowStrainRate\\ 
 &= \Omega^*(T)\ehat{3}+[\omega\Omega_f(T) + \Omega(T)] \ehat{3} - \tilde{\tensor{h}} \flowStrainRate. \label{gvt2}
    \end{align}
\end{subequations}
Noting the assumption of planarity used in the above,  we also require $\tilde{\tensor{h}} \flowStrainRate$ to be parallel to $\ehat{3}=\e_3$. While not assured for general particle shapes, this is guaranteed by only relatively weak symmetry constraints on the swimmer (which must apply throughout its entire deformation), such as those detailed in \cref{sgedgc,app: shapes}. Finally, while rotation out of the plane of flow is not admissible, translation perpendicular to the plane of flow (with the symmetry broken by the swimmer shape) can be accommodated, so that we do not a priori require that $ -\tilde{\tensor{g}}\flowStrainRate$ has no component perpendicular to the flow plane. 
In \cref{sgedgc}, we simplify $- \tilde{\tensor{h}} \flowStrainRate$ and  $- \tilde{\tensor{g}} \flowStrainRate $ and proceed to explore the resulting equations of motion. We note that in the main text (below \cref{{UcI}}) we note that inefficiency forces a further constraint, in particular that the combination of fast rotational oscillations and fast (but zero-average) translational dynamics do not interact to produce a net fast swimming speed. 

\section{Admissible swimmer shapes}\label{app: shapes}
Here, we describe various classes of swimmer shapes that are admissible within the framework of this manuscript.
\subsection{$C_{ nv}$ bodies}
These are swimmer shapes that, for all times of the gait cycle, possess a helicoidal symmetry of degree $n\geq 3$ along with $n$ reflection planes containing the helicoidal axis. For such bodies, we have
\begin{equation}
    \eta_2=0, \quad \lambda_2\equiv-B, \quad \lambda_5,\, \eta_3,\, \eta_4 \neq 0
\end{equation}
for $n=3$, while for $n\geq4$ we have
\begin{equation}
    \lambda_5=\eta_2= 0, \quad\lambda_2\equiv -B, \quad \eta_3,\,\eta_4 \neq 0.
\end{equation}

\subsection{$C_{nh}$ bodies}
These swimmers are a subset of $C_{nv}$ bodies and possess an additional reflection symmetry perpendicular to the helicoidal axis, i.e. a fore-aft symmetry. With this, we have
\begin{equation}
    \Omega_f(T)=\Omega(T) = \lambda_5=\eta_2=\eta_3=0, \quad \lambda_2\equiv -B, \quad \eta_4 \neq 0
\end{equation}
for $n=3$, while for $n\geq4$ we have
\begin{equation}
    \Omega_f(T)=\Omega(T) =\lambda_5=\eta_2=\eta_3=\eta_4= 0, \quad \lambda_2\equiv -B \neq 0.
\end{equation}
Note that the fore-aft symmetry of such swimmers also entails that the angular velocity in the absence of flow must be zero, as this would otherwise break fore-aft symmetry. Hence, we also have $ \Omega_f(T)=\Omega(T) =0$ for these swimmers. 
 
\subsection{$D_{n}$ bodies}
This class of bodies includes swimmer shapes that, for all times of the gait cycle, possess a helicoidal symmetry of degree $n\geq 4$. In addition, they possess dihedral symmetry associated with $n$-axes perpendicular to the helicoidal axis, around which a rotation of $\pi$ leaves the body invariant. Then, we have 
\begin{equation}
   \lambda_5=\eta_3=0, \quad \lambda_2\equiv- B, \quad \eta_2, \eta_4 \neq 0. 
\end{equation}

\subsection{$D_{nh}$ bodies} 
These swimmers are a subset $D_n$ bodies and additionally possess a reflection symmetry perpendicular to the helicoidal axis, that is a fore-aft symmetry. Then we have 
\begin{equation}
    \Omega_f(T)=\Omega(T)  = \lambda_5=\eta_2=\eta_3=0, \quad \lambda_2\equiv -B, \eta_4 \neq 0.
\end{equation}

\subsection{Bodies of revolution}
Swimmers that are bodies of revolution satisfy the symmetries of both the $D_n$ and $C_{ nv}$ ($n\geq 4$) bodies and, when possessing additional fore-aft symmetry, the symmetries of both the $D_{nh}$ and $C_{nh}$ ($n\geq 4$) bodies. In general, however, there are no further simplifications. 

Thus, we have the translational equation of motion
\begin{multline}\label{xeqnf1} 
\diff{\x_c}{t} = \flowVel_c + \omega U(T) \ehat{1}+ P\ehat{1} -\eta_2 (T)\hat{E}^*_{12}(T,\sin2\theta,\cos2\theta)\hat{\e}_3 \\ 
+\eta_3 (T)\vec{B}_0(T,\sin2\theta,\cos2\theta)\hat{\e}_1-\eta_4 (T)\hat{E}^*_{22}(T,\sin2\theta,\cos2\theta)\hat{\e}_2, 
\end{multline}  
with 
\begin{equation}
     \vec{B}_0(T,\sin2\theta,\cos2\theta)=
    \flowStrainRate(T) - [E^{11}(T)\cos2\theta+E^{12}(T)\sin2\theta]\mat{I}
\end{equation}
in this case. Analogously, the angular equation of motion simplifies to 
\begin{multline}\label{eq: intermediate orientation2}
 \diff{\theta}{t} = \Omega^*(T)+\omega \Omega_f(T)+\Omega(T) 
 +[\lambda_5(T)E^{12}(T)-B(T)E^{11}(T)]\sin{2\theta} \\  
 +[\lambda_5(T)E^{11}(T)+B(T)E^{12}(T)]\cos{2\theta}.
\end{multline}

\section{Symmetry simplifications of the leading order multiscale equations of motion}\label{SymmSimp}
Here, we consider further simplifications to the leading order multiscale equations of motion given by \cref{teqnf,eqnxc00} once the swimmer possesses additional symmetries or does not have a rapid oscillatory motion in a quiescent fluid.

\subsection{Fore-aft symmetry and the absence of fast rotation}
Suppose that there is no self-induced rapid oscillatory rotation ($\Omega_f=0$), so that the swimmer rotation rate in a quiescent fluid reduces to
\begin{equation}
   \vec{\Omega}(T,\vec{0})=(\omega\Omega_f(T) + \Omega(T)) \e_3 = \Omega(T)\e_3.
\end{equation}
Then, we have $\Omega_{fI}=0=\Psi$ and, hence, for the angular equations 
we have 
\begin{equation}
  \theta_0(t,T)=\avg{\theta}_{0}(t).  
\end{equation}
As a result, the angular dynamics no longer possesses a fast oscillation around the averaged swimmer orientation associated with $\avg{\theta}_{0}(t)$, which reduces the complexity of the rotational and translational equations substantially. 

Firstly, \cref{cstarstar} reduces to 
\begin{subequations}\label{bcsimp}
 \begin{align}
 b_{**}(T) &=  B(T)E^{11}(T)-\lambda_5(T)E^{12}(T)=b_*(T), \\ 
c_{**}(T) &=   -B(T)E^{12}(T)-\lambda_5(T)E^{11}(T) =c_*(T). 
 \end{align} 
\end{subequations}
Hence, the leading order angular equation  takes the same form as before, i.e.
\begin{equation} \label{teqnfapp} 
    \diff{\avg{\theta}_0}{t} = a - b \sin 2\avg{\theta}_0 - c \cos2\avg{\theta}_0, \quad a=\avg{a}_*,\, b=\avg{b}_{**} ,\, c= \avg{c}_{**},
\end{equation} 
where the fast-time averages $b=\avg{b}_{**}$ and $c=\avg{c}_{**}$ that classify the rotational equations of motion are simplified significantly according to \cref{bcsimp}. Nonetheless, the interpretation that they represent measures of the rate of strain of the background flow, modulated by swimmer properties, is still retained. 

Considering the translational equation \cref{eqnxc00}, with the definition
\begin{equation}
    U_I = \int_0^T U(S)\intd{S},
\end{equation}
and noting that $\avg{U}_I=0$ by inheritance of the periodicity of $U(T)$, 
we have the further simplifications that 
\begin{gather*}
    U_{sI}=0, \quad \Phi_s(T)=0, \quad U_I(T)=U_{cI}(T), \quad\avg{U}_{cI}=0, \\
    \Phi_c(T) = U_I(T), \quad \chi(T) = a_*(T) - b_*(T)\sin2\avg{\theta}_0-c_*(T)\cos2\avg{\theta}_0. 
\end{gather*}
Noting once more that $\avg{\flowVel_{tr}}=\vec{0}$ without loss of generality by choice of the inertial reference frame, we have that \cref{zeqn} reduces to 
\begin{equation}\label{zeqnapp} 
\diff{\avg{z}_0}{t} = -\avg{\eta_2 (T)E^{12}(T) }\cos(2\avg{\theta}_0) + \avg{\eta_2 (T)\hat{E}^{11}(T)}\sin(2\avg{\theta}_0), 
\end{equation}
while \cref{eqnxc000} reduces to 
\begin{align}
\begin{split}
\diff{\avg{\x}_0}{t} &=
    \left[ \mat{A}\avg{\x}_{0} +\avg{U_I\mat{\Lambda}} +P \mat{I} +\avg{\eta_3(T)\vec{B}_0(T,\sin2\avg{\theta}_0,\cos2\avg{\theta}_0)}\right]\ehat{10}(\avg{\theta}_0) \\  
    & \quad - \left[\avg{\eta_4 \hat{E}^*_{22}  (T,\sin2\avg{\theta}_0,\cos\avg{\theta}_0)}+\avg{\chi U_{I} }\right]\ehat{20}(\avg{\theta}_0)
    \end{split}
\\\begin{split}&= 
\begin{pmatrix}
\avg{E^{11}} & \avg{E^{12} -\Omega^*}
\\
\avg{E^{12}+\Omega^*} & -\avg{E^{11} }
\end{pmatrix} \avg{\x}_{0} 
 \\ &\quad+ 
 \left[(\avg{\eta_4E^{11}} + \avg{U_I c_*} ) \cos2\avg{\theta}_{0} +
 (\avg{\eta_4E^{12}}+ \avg{U_I b_*} ) \sin2\avg{\theta}_{0}-\avg{U_I a_*}\, \right]\ehat{20}(\avg{\theta}_0) \\ 
  &\quad+ \left[P -\avg{\eta_3E^{11}}\cos2\avg{\theta}_{0}-\avg{\eta_3E^{11}}\cos2\avg{\theta}_{0}\right]\ehat{10}(\avg{\theta}_0)  
\\ &\quad+ 
\begin{pmatrix}
\avg{U_I E^{11}}+\avg{\eta_3 E^{11}} & \avg{U_I ( E^{12} -\Omega^*)} +\avg{\eta_3 E^{12}} \\ 
\avg{U_I ( E^{12} +\Omega^*)} +\avg{\eta_3 E^{12}} &  -\avg{U_I E^{11}} -\avg{\eta_3 E^{11}}
\end{pmatrix}\ehat{10}(\avg{\theta}_0).  \label{eqnxc000app}
\end{split}
\end{align}

\subsection{The equations of motion for a body of revolution with fore-aft symmetry}

For a body of revolution with fore-aft symmetry we additionally have 
\begin{equation}
    \lambda_5 = \Omega(T)=\eta_2=\eta_3=\eta_4=0,
\end{equation}
so that the influence of the rate of strain is only through the Bretherton parameter $B(T)\equiv-\lambda_2(T)$. Then, 
with $\avg{\flowVel_{tr}}=\vec{0}$ by the choice of the inertial reference frame, the translational equations of motion further simplify to
\begin{subequations}\label{eqnxc000app1}
\begin{align}
\diff{\avg{z}_0}{t}&=0,  \\  
\diff{\avg{\x}_{0}}{t}
&=
 \left[ \mat{A}\avg{\x}_{0} 
+\avg{U_I\mat{\Lambda}}  
+P \mat{I} \right]\ehat{10}(\avg{\theta}_0)- 
\avg{\chi U_{I} } \ehat{20}(\avg{\theta}_0)
\\ 
\begin{split}
    &= 
\mat{A} \avg{\x}_{0}  -
( \avg{U_I a_*}-
  \avg{U_I b_*}  \sin2\avg{\theta}_{0}- \avg{U_I c_*}  \cos2\avg{\theta}_{0} )\ehat{20}(\avg{\theta}_{0}) \\  &\quad+ \begin{pmatrix}
P+\avg{U_I E^{11}} & \avg{U_I ( E^{12} -\Omega^*)} \\
\avg{U_I ( E^{12} +\Omega^*)}   & P -\avg{U_I E^{11}} 
\end{pmatrix}\ehat{10}(\avg{\theta}_{0}).
\end{split}
\end{align} 
\end{subequations}

\bibliographystyle{jfm.bst}
\bibliography{library.bib}
\end{document}